\documentclass[dvipsnames, 12pt]{article}
\usepackage[margin=30mm]{geometry}
\usepackage{array}

\usepackage[sort,comma,authoryear]{natbib}
\usepackage{authblk}
\usepackage{caption}
\usepackage{framed}
\usepackage{xr-hyper}
\usepackage{hyperref}
\hypersetup{
    colorlinks=true,
    linkcolor=cyan,
    filecolor=cyan,      
    urlcolor=cyan,
    citecolor=cyan,
    breaklinks=true}
\usepackage{color,soul}
\usepackage{graphicx}
\usepackage{amsmath}
\usepackage{amssymb }
\usepackage{setspace}
\usepackage{threeparttable}
\usepackage{booktabs}
\usepackage{subcaption}
\usepackage{diagbox}
\usepackage{bm}
\usepackage{algorithm}
\usepackage{algpseudocode}

\title{Fake News Detection via\\ Wisdom of Synthetic \& Representative Crowds}

\author[a]{François t'Serstevens \thanks{Corresponding Author: François t'Serstevens. f.tserstevens@maastrichtuniversity.nl}} 
\author[b]{Roberto Cerina}
\author[c]{Giulia Piccillo}

\affil[a]{\footnotesize Department of Data Analytics and Digitilisation, Maastricht University, 53 Tongersetraat, 6211 LM Maastricht, Netherlands}

\affil[b]{Department of Media Studies, University of Amsterdam, 1090 GN Amsterdam Turfdraagsterpad 9}

\affil[c]{\footnotesize Department of Economics, Maastricht University, 53 Tongersetraat, 6211 LM Maastricht, Netherlands}

\begin{document}
\date{\today}


\newcommand*{\addFileDependency}[1]{
\typeout{(#1)}
%
%

%
\IfFileExists{#1}{}{\typeout{No file #1.}}
}\makeatother

\newcommand*{\myexternaldocument}[1]{%
\externaldocument{#1}%
\addFileDependency{#1.tex}%
\addFileDependency{#1.aux}%
}

\myexternaldocument{SI}

\maketitle

\pagebreak

\begin{abstract}
Social media companies have struggled to provide a democratically legitimate definition of ``Fake News''. Reliance on expert judgment has attracted criticism due to a general trust deficit and political polarisation. Approaches reliant on the ``wisdom of the crowds'' are a cost-effective, transparent and inclusive alternative. This paper provides a novel end-to-end methodology to detect fake news on $\mathbb{X}$ via ``wisdom of the synthetic \& representative crowds''. We deploy an online survey on the Lucid platform to gather veracity assessments for a number of pandemic-related tweets from crowd-workers. Borrowing from the MrP literature, we train a Hierarchical Bayesian model to predict the veracity of each tweet from the perspective of different personae from the population of interest.
We then weight the predicted veracity assessments according to a representative stratification frame, such that decisions about ``fake'' tweets are representative of the overall polity of interest. Based on these aggregated scores, we analyse a corpus of tweets and perform a second MrP to generate state-level estimates of the number of people who share fake news. We find small but statistically meaningful heterogeneity in fake news sharing across US states. At the individual-level: i. sharing fake news is generally rare, with an average sharing probability interval $[0.07,0.14]$; ii. strong evidence that Democrats share less fake news, accounting for a reduction in the sharing odds of $[57.3\%,3.9\%]$ relative to the average user; iii. when Republican definitions of fake news are used, it is the latter who show a decrease in the propensity to share fake news worth $[50.8\%, 2.0\%]$; iv. some evidence that women share less fake news than men, an effect worth a $[29.5\%,4.9\%]$ decrease.

\end{abstract}

\newpage

\section{Introduction}

Fake news detection has historically relied on fact-checkers to objectively investigate the veracity of claims. Bias-related concerns 
have prompted a critical reevaluation 
\citep{nieminen2019fighting, poynter2019fact, rich2020research, walker2019republicans}. 
Claims about the veracity of a piece of digital content (e.g. a tweet, a post, a news article, etc.) need to have legitimacy in the eyes of online users, in order to, in turn, legitimise content moderation policies \citep{pan2022comparing}. This paper provides a methodology to endow fake news detection with \emph{democratic legitimacy} -- this is in contrast with other forms of legitimacy, such as those provided by \emph{deference to expertise}, or the (perceived) \emph{independence} of content-detection algorithms \citep{pan2022comparing}. 


We contribute to the methodological literature on the use of wisdom-of-the-crowd for the veracity assessment of digital content. Our first contribution relates to the estimation and computation of optimal crowd assessments of veracity. We propose two estimation methods: \emph{naive} and \emph{model-based}. Naive estimation takes the raw veracity assessments from the sample, without correcting for the underlying unrepresentativeness of crowd-workers. Model-based estimation instead leverages a Hierarchical Bayesian model to learn from the crowd-workers, and generate predicted veracity scores for a series of personae, taking into account the attributes of the persona, the tweet and their interactions. The personae's veracity scores are then aggregated, 
recovering a robust assessment in accordance to the wisdom of the synthetic personae.

We further contribute to this literature by introducing four aggregation methods for the crowd's assessments. The aggregation methods are designed to weigh the veracity assessments provided by survey participants in a manner that reflects the broader sentiments of the crowd, either as a whole (e.g. relative to the true preferences of the whole country under the ``one-man, one-vote'' principle) or within specific population segments (e.g. relative to the true preferences of Republicans, as opposed to Democrats, or the young, as opposed to the old, etc.). 

Our wisdom-of-the-crowd metrics are explainable, approximately representative, and provide a fruitful avenue to keep the ``human in the loop'' \citep{link2016human} for content moderation on social media.\\

\noindent We further produce a substantive contribution to the study of the individual- and area-level correlates of fake news sharers across the United States. We fit a second Hierarchical Bayesian model, taking the optimal veracity assessment described above as the dependent variable, and a set of inferred characteristics from social-media users who have shared these tweets as the set of covariates. At the individual-level, we find that fake news sharing is generally rare. Consistent with previous research \citep{guess2019less} relying on fact-checkers, we find that Democrats share less fake news when compared to the average $\mathbb{X}$ user. This trend persists even when our aggregation approach is tuned to mimic a jury composed of equal parts of Republicans and Democrats. Importantly, when the assessment is conducted exclusively by Republicans, we find that Republicans exhibit a lower propensity to share fake news. This suggests fundamentally different understandings of the fake news phenomenon across partisan lines. We also find some evidence to the effect that women share less fake news online than men. 

A Multilevel Regression and Post-Stratification (MrP) \citep{park2004bayesian} approach is then used to generate state-level estimates of the average propensity to share fake news. 
There is disagreement about the ``size'' of the fake news problem \citep{lyons2020bad}. It is often difficult to measure this coherently. Our proposal is to use estimates of the measurable concept \emph{``probability to share news which is deemed fake by the crowd''} to provide policy-makers with a realistic picture. Across accuracy metrics, the state-wide fake news sharing estimates maintain relatively consistent rankings, with states like Tennessee and DC consistently showing a higher and lower propensity for sharing fake news respectively. The heterogeneity across states is small but statistically meaningful.\\

The rest of the paper is structured as follows, Section \ref{lit} introduces the challenges in identifying fake news and making representative inferences. Section \ref{data} describes the survey setup, detailing the collection of data from crowd-workers and the integration of external datasets. Section \ref{sc:accuracy} details the calculation and use cases of a number of candidate veracity scores. Section \ref{map} builds upon the calculated tweet-accuracy metrics to estimate the spread of fake news across the United States. We discuss our findings in Section \ref{discussion}.

\section{Identifying Fake News on Social Media}\label{lit}



\citep{lazer2018science} define fake news as ``\emph{fabricated information that mimics news media content in form but not in organizational process or intent}''. On social media, process and intent are largely unobservable. \citep{lazer2018science}'s definition is not immediately helpful to inform the discernment of fake news on social media, where decisions on content moderation are typically based exclusively on the observable output of a given process -- i.e. the content of a post or a tweet. To overcome this hurdle, three primary modes of detection have been employed by industry practitioners and academics: i. \emph{expert assessment}; ii. \emph{crowd-sourced assessment}; iii. \emph{computational methods}.


\textbf{Expert Assessment}. Fact-checking by professional journalists has been widely adopted by social media companies, as evidenced by Facebook, Instagram and $\mathbb{X}$ (formerly Twitter) all implementing some form of fact-checking oversight at some point in recent years \citep{insta2019comb, facebook2023how, x2023how}. Academics have also resorted to professional fact-checkers to identify fake news \citep{allcott2017social, mena2020cleaning, pennycook2019fighting, vosoughi2018spread}. A finding in this literature is that Republicans tend to be significantly more likely to spread fake news than Democrats \citep{guess2019less}. Republicans generally have low-trust in fact-checkers, and contest the validity and legitimacy of their assessments \citep{nieminen2019fighting}. Republicans are generally under-represented in the journalistic profession \citep{willnat2022american}, and tend to have a large trust-gap when it comes to journalism and media, relative to Democrats \citep{liedke2022us}. 

\textbf{Crowd-sourced Assessment}. Scholars and practitioners have recently explored the wisdom of the crowds to identify fake news \citep{pennycook2019fighting}. This crowd-based definition of the truth comes with two main advantages over fact-checking, i. a universal, understandable, measurable and verifiable definition of the truth (i.e. ``that which the majority of the crowd deems to be true''); ii. superior scalability.
Provided an initially representative sample, crowd-sourced assessments are significantly correlated to fact-checkers' ratings, even when a limited number of reviews per claim are available \citep{allen2021scaling}. 
The difficulty of crowd-sourced ratings lies in the representativeness of the initial reviewer sample, i.e. a Democrat-leaning pool of laypeople would favour a Liberal narrative over a Conservative one, in their veracity assessments. 

\textbf{Computational Methods}. Enabled by the data-rich environment of social media, computational methods leverage large amounts of available information to train classifiers for automated fake content detection. The features included in these classifiers are varied, often including: i. content features such as the source of the tweet or post and its headline; ii. social and personal context data, such as the parameters of users' networks, the self-reported content of users' profiles, etc.; iii. various meta-data, including account creation dates, posting frequency, etc.; iv. various linguistic features \citep{shu2017fake}. Computational models have shown promise to assess fake news automatically with a handful of models coming close to human assessment \citep{agarwal2019analysis, hussain2020detection}. 
It is worth noting that all machine learning models require an exogenous definition of the truth, either crowdsourced or from expert review, which these models learn to approximate.

Notably, the methods outlined above are not mutually exclusive; $\mathbb{X}$'s recently implemented ``Community Notes'' leverages both a machine learning algorithm and user input \citep{x2023com}. Likewise, user flags have been used to select review-worthy claims in an effective manner \citep{tschiatschek2018fake}.

\subsection{Online Sharing Dynamics}

Across social media fake news consumption is highly heterogeneous, with a handful of individuals primarily responsible for posting and sharing fake news \citep{grinberg2019fake, guess2019less}. There is some evidence that echo chamber structures and bots have fostered the spread of fake news \citep{difonzo2011echo}, though the extent of these effects is disputed \citep{cinelli2021echo, guess2018avoiding, vosoughi2018spread}. Behavioural elements, such as analytical thinking and perceived accuracy or ``fear of missing out'' have been identified as consistent factors in the sharing process \citep{pennycook2019lazy, talwar2019people, t2022fake}. Although behavioural elements appear central to fake news sharing, they are not typically directly measurable on social media \citep{osmundsen2021partisan}.
Socio-demographic factors have been shown to be correlated with fake news sharing in the context of the $2016$ US Presidential election, with Republicans in older age groups consistently sharing more fake news than their Democrat counterparts according to fact-checkes' data \citep{guess2019less, grinberg2019fake}. 
Unlike behavioural elements, age, gender, and political alignment of users can often be observed directly in the information-rich social media environment. 

\section{Data Collection} \label{data}

This section covers the gathering of a pool of tweets; the setup of the survey experiment, as well as the derivation of user-characteristics from social media. 
The survey design, data (excluding sensitive/identifiable information) and the \texttt{R} \citep{r2013r}, \texttt{python} \citep{van1995python} and \texttt{Stan} \citep{carpenter2017stan} code used for the analyses are available online through \texttt{GitHub}\footnote{\url{https://github.com/ftserstevens/mapping}}. The survey experiment was approved by the Maastricht University ethics committee, with review reference number ERCIC\_348\_15\_04\_2022.

\subsection{Tweet Selection}

A total of $247,725$ tweets, posted from  May $2022$ to December $2022$, were downloaded from the $\mathbb{X}$ API using a list of COVID-19 related keywords\footnote{\textit{Corona, COVID, COVID-19, Coronavirus, Facemasks, Vaccine}}. U.S. Geotagged tweets were exclusively selected as reliable localisation data was necessary to enable post-stratification at the state level. Therefore, $\mathbb{X}$ users who opted out of the geolocalisation feature were excluded from the analysis a priori.
From this initial pool of tweets, a final set of $5,513$ was selected for review. Although this represents a considerable downsizing, reducing the tweet pool enabled the crowd-sourced approach. 
The reasoning and mechanics of this selection process are as follows:

\begin{enumerate}

\item \textbf{Non-Relevant Tweets}. A majority of downloaded tweets did not contain any verifiable information that either laypeople or fact-checkers could review. Consider the following tweet: 
``Vaxxed and ready to tackle life with extra protection! Grateful for science and a brighter future.''. It does not contain any verifiable information nor is it relevant to the societal debate on fake news. 
Most tweets found on $\mathbb{X}$ were akin to this example. $\mathbb{X}$'s built-in context annotations\footnote{At the time $\mathbb{X}$ provided the option to download context annotations through its API. These context annotations are used by $\mathbb{X}$ to create its trending topics. They denote whether a tweet is about one or multiple personalities, events, objects, etc. Section \ref{sc:external data} details their use in this paper.} were used to select tweets that related to both COVID-19 and political events. Though the filtering did not fully exclude non-informative tweets, it drastically reduced their prevalence in the sample.

\item  \textbf{Effective Usage of Resources}. An equilibrium between the number of tweets in the review pool and the average amount of reviews per tweet needed to be determined to ensure efficient usage of the finite amount of veracity assessments. On the one hand, the greater the number of tweets in our pool the greater the diversity of $\mathbb{X}$ users who shared these tweets in the sample. On the other hand, the fewer tweets in the final pool the more crowd-sourced reviews we could afford per tweet. Preceding studies have suggested that there were diminishing returns after $5$ reviews per tweet \citep{allen2021scaling}.

\item \textbf{Optimal Number of Reviews per Tweet}.
Given the importance of the first $5$ reviews on a given tweet, the optimal number of reviews per tweet was estimated to maximise the number of tweets with more than $5$ reviews.
Given a finite number of available reviews ($\sim 40,000$ according to our survey setup and resources constraints), an average of $6$ to $7$ reviews per tweet would maximize the number of effective tweets, i.e. tweets with more than $5$ reviews.
This translates into an initial sample of approximately $6,500$ to $5,500$ tweets. 
Consequently we focused on tweets posted from May $2022$ onwards. Together, this time-bound filter and the aforementioned relevance filter resulted in a final pool of $5,513$ tweets. An average of $7.73$ reviews per tweet were gathered from the survey respondents for all $5,513$ tweets. The number of reviews per tweet here is higher due to a surplus of survey participants.
\end{enumerate}

\subsection{Estimating $\mathbb{X}$ Users’ Socio-Demographics}
\label{sc:external data}

In addition to the data provided through the $\mathbb{X}$ API, author information was supplemented with estimates of author gender, age and political alignment. 
Author age and sex were determined through the \texttt{m3} inference pipeline \citep{wang2019demographic}, which leverages the author's profile picture and description to generate a probabilistic estimate of age category and gender. A majority of the sample ($57\%$) was classified as $40+$ years old by the algorithm. The remaining percentages are spread across the $18-30$ and $30-40$ categories equally. 
Figure \ref{fig:matchplot_census} depicts the proportions of users within the sample and how they differ from the population. The inferred marginal distributions of socio-demographic attributes match that of the census surprisingly closely.

Political affiliation was estimated through the elite misinformation-exposure estimation tool \citep{mosleh2022measuring}. The tool identifies elite accounts within the networks of $\mathbb{X}$ users. Elite accounts are typically well-known $\mathbb{X}$ profiles whose partisanship was scored by PolitiFact. The elite scores are aggregated per user to estimate the political affiliation of the user \citep{barbera2015birds}. Users above below or in between $-.25$ and $.25$ were classified respectively as either Democrats, Republicans or Neutrals. Users who did not follow any elite accounts were also classified as Neutrals.
Across all authors, $36\%$ were Democrat-leaning, $17\%$ had neutral scores and $21\%$ were Republican-leaning, the remaining $26\%$ did not follow any elite accounts. The socio-demographic characteristics of the tweet authors and their differences with the census data are summarized in figure \ref{fig:matchplot_twitter} -- note that ground-truth for partisanship scores comes from the American National Election Study, rather than the census.

Information on tweet content and topic was derived through $\mathbb{X}$'s native context annotations. $\mathbb{X}$ uses context annotations to classify tweets in one or multiple topics. For instance, the annotations are used to identify the trending topics on $\mathbb{X}$.
Only the $30$ most mentioned context annotations were recorded. A given tweet could include several annotations at once, e.g. a tweet might address both ``Governmental institutions'' and ``Joe Biden''  simultaneously. Highly concurrent annotations (over $60\%$ co-occurrence both ways) were excluded.

\subsection{Crowd-Workers Socio-Demographics}

\begin{figure}
    \centering
    \begin{subfigure}{0.48\textwidth}
        \centering
        \includegraphics[width=1\linewidth]{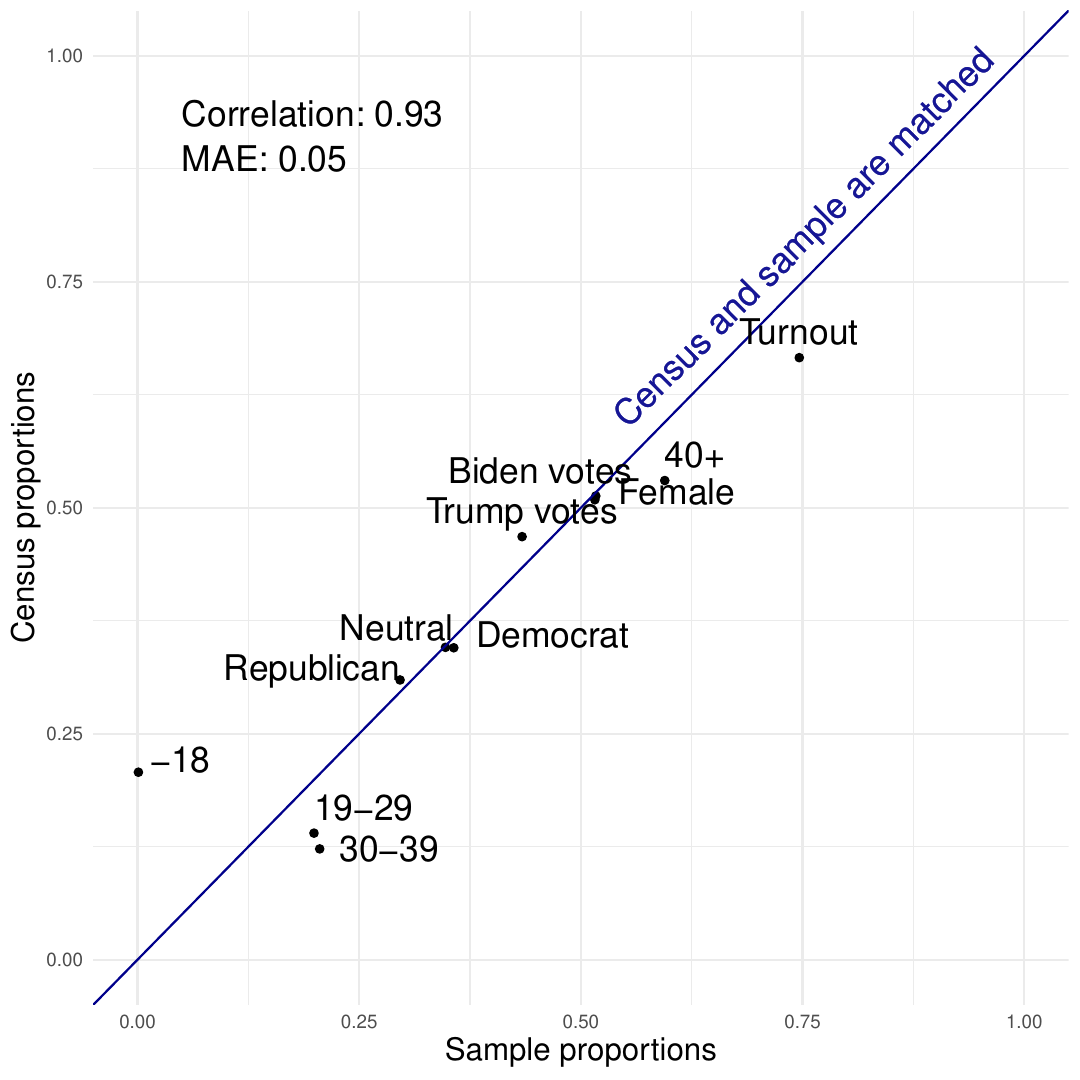}
        \caption{Lucid Raters}
        \label{fig:matchplot_census}
    \end{subfigure}
    \hfill
    \begin{subfigure}{0.48\textwidth}
        \centering
        \includegraphics[width=1\linewidth]{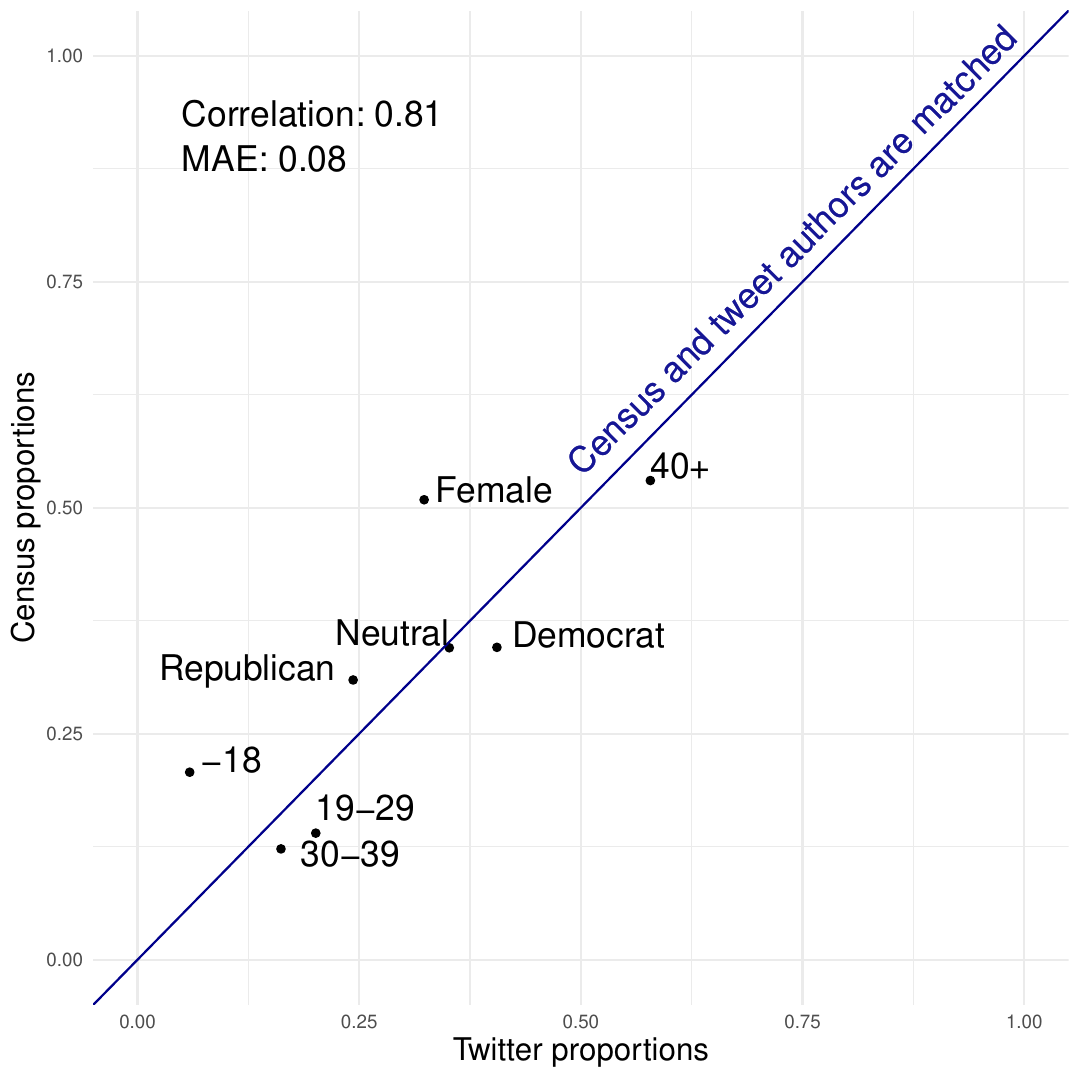}
        \caption{X Users}
        \label{fig:matchplot_twitter}
    \end{subfigure}
    \caption{The y-axis represents the census / ground-truth proportions and the x-axis represents the Lucid and $\mathbb{X}$ sample proportions. Any point on the blue line represents a matching proportion. Points that are to the left/right of the blue line are under-/over- represented.}
    \label{fig:matchplot}
\end{figure}

Between January and March $2023$, we recruited $5,154$ U.S.-based participants from Lucid. 
From the initial pool of participants, $2,196$ participants passed the attention checks, and $166$ participants did not complete the survey fully. A total of $2,030$ unique participants were finally analysed. The survey demographics and their difference with the census data are summarized in figure \ref{fig:matchplot}. 
The survey was designed with quotas ensuring equal representation of both genders, representative age groups, and a maximum limit of $100$ participants from each state.
Every ZIP code in the US was represented at most once in the sample, participants from an already represented ZIP code were rejected. This feature was implemented to ensure a diverse sample population, it enforces a more geographically diverse sample suited to the subsequent implementation of MrP. On the whole, both our sample of Lucid raters and -- surprisingly -- our sample of $\mathbb{X}$ users were reasonably representative of the population as a whole, at the level of the marginal distributions described in the plots.

\subsection{Veracity Assessment Protocol}

Each survey participant assessed the accuracy of $20$ tweets. The tweets were selected randomly from the aforementioned curated pool and were shown one at a time. For each tweet, laypeople indicated their perceived accuracy using a $4-$point accuracy scale akin to previous research: \textit{Not at all accurate, Not very accurate, Somewhat accurate and Very accurate} \citep{pennycook2019fighting}. 
As the analyses require a numerical quantification of the perceived accuracy, the aforementioned accuracy scale was converted to a numerical scale from 1 to 4, 1 being \textit{Not accurate at all} and 4 being \textit{Very accurate}.

Three attention checks were placed in the assessment phase. Participants who failed more than one were excluded from the survey. Attention checks prompted laypeople to select pre-defined accuracy and confidence ratings, e.g., ``Very accurate'' and ``Very confident''. The attention checks had the same format as the tweet accuracy rating tasks. 


\section{Aggregating Crowd Assessments}
\label{sc:accuracy}

To estimate the veracity of each tweet according to the wisdom of the crowd, this paper introduces a total of eight veracity metrics, representing different choices in the \emph{estimation} and \emph{aggregation} phases.

Two \emph{estimation} strategies are presented: \emph{naive} and \emph{model-based}. Naive veracity scores do not use participant information such as their political party, whilst model-based estimates do. The necessity of the model-based estimates stems from the non-representativeness of the Lucid sample. The model-based metrics aim to estimate each tweet's veracity as if a fully representative crowd assessed it.

For both naive and model-based estimation, four \emph{aggregation} strategies are employed: i. \textit{sample}-based aggregation, which simply takes the average of the observed veracity assessments as the final veracity estimate, for each tweet; ii. \textit{balanced} aggregation, which uses a bootstrap procedure to ensure an equal amount of Democrat and Republican raters assess each tweet; iii. \textit{population} aggregation, where a representative stratification frame is used weight the assessments according to the demographic characteristics of the raters; iv. \textit{partisan} aggregation, which represents the average veracity assessment from raters who identify with a given political party. 
Table \ref{tab:accmetrics} summarizes the main veracity metrics. This section details the estimation and aggregation process, and assesses the similarity of these metrics.

\begin{table}[htbp]
  \centering
  \begin{tabular}{p{3.75cm}|p{5cm}p{5cm}}
  \addlinespace
  \addlinespace
    \toprule
    \diagbox[width=4cm]{\small{\textbf{Aggregation}}}{\small{  
 \textbf{Estimation}}}& \multicolumn{1}{c}{\emph{naive}} & \multicolumn{1}{c}{\emph{model-based}} \\
    \midrule
    \emph{sample} & simple average of observed assessments & simple average of fitted values \\
    \addlinespace
    \emph{balanced} & weighted average where weights enforce equal number of observed assessments from Rs and Ds & weighted average where weights enforce equal number of fitted values from Rs and Ds \\
    \addlinespace
    \emph{population} & weighted average where weights are derived via Iterative Proportional Fitting to match census margins & fully post-stratified estimate of predicted values  \\
    \addlinespace
    \emph{partisan} & observed assessments from Rs or Ds & post-stratified predicted value for Rs or Ds, where the stratification frame is representative of the true R or D population.\\
    \addlinespace
    \bottomrule
  \end{tabular}
\caption{Summary of the veracity assessment metrics according to their respective estimation and aggregation procedures.}  
\label{tab:accmetrics}

\end{table}

\subsection{Naive Estimation}
Take $y_{ti} \in \{1,...,4\},\mbox{ } \forall : \mbox{ } i \in \{1,...,N\}, \mbox{ } t \in \{1,...,M\}$ to be the veracity assessment provided by a Lucid crowd-worker $i$, for tweet $t$. Note that not every tweet is labeled by every worker, so $\bm{\iota}_t \in \{\bm{\iota}_1,...,\bm{\iota}_{M}\}$ is the subset of $n_t$ workers that produced assessment specifically for tweet $t$.

The first naive estimation method calculates the veracity score per tweet by taking the simple average of the veracity assessments in our sample of crowd-workers. This measure is hereafter referred to as $\nu^{ \{ \mbox{\scriptsize naive}, \mbox{ \scriptsize sample}\} }$, its formal definition can be found in Equation \ref{eq:raw_accuracy}. In an ideal scenario with a sufficiently large and diverse group of reviewers for each tweet, and a representative sample, the characteristics of the reviewers would accurately reflect those of the broader population for all tweets, and hence this metric should recover an unbiased measure of the wisdom-of-the-crowd veraicty score estimate. In the experiment at hand, an average of $7.73$ reviews per tweet were made reviews, making it is unlikely that the reviewer pool of every tweet would represent the views of the overall population.
\begin{equation}
\label{eq:raw_accuracy}
\nu^{ \{ \mbox{\scriptsize naive}, \mbox{ \scriptsize sample}\} }_{t} = \frac{1}{n_t} \sum_{i \in \bm{\iota}_t} y_{ti}
\end{equation}

\noindent The second metric we present is the balanced naive veracity estimate, $\nu^{ \{ \mbox{\scriptsize naive}, \mbox{ \scriptsize balanced}\} }$, which ensures that an equal number of Democrat and Republican assessors are considered for each tweet. Democrat and Republican groups are defined here by their recalled voting in the $2020$ election. The metric is formally defined in Equation \ref{eq:eqw_accuracy} \footnote{Though $\nu^{ \{ \mbox{\scriptsize naive}, \mbox{ \scriptsize balanced}\} }$ enforces a political balance, it is worth noting that both the independent- and non-voters were excluded from the aggregation altogether. }. In line with previous literature \citep{allen2021scaling, pennycook2019fighting}, we implement a bootstrapping procedure (pseudo-Algorithm \ref{algo:boot}) to deal with workers in the over-represented political group, for a given tweet. For each bootstrap iteration, an equal number of assessments from both Democrats and Republicans are sampled. The assessments from the over-represented political group were resampled at each iteration to match the number of the under-represented group. This enables the use of all the reviews within the over-represented category. An average of all the bootstraps is then used to generate the final balanced estimate.
\begin{equation}
\label{eq:eqw_accuracy}
\nu^{ \{ \mbox{\scriptsize naive}, \mbox{ \scriptsize balanced}\} }_{t} = \frac{1}{2} \{ \bar{\bm{y}}^D_t + \bar{\bm{y}}^R_t \}
\end{equation}
\begin{algorithm}
\caption{Balancing assessments across groups via Bootstrap.}
\label{algo:boot}
\begin{algorithmic}[1]
\For {$t \in \{1,...,M\}$} \Comment{For each tweet}
    \State $n^j_{t} = \|\bm{y}^{j}_{t}\| \quad \forall \, j \in \{R, D\};$  \Comment{Count n. assessments across groups}
    \If {$n^j_{t} > n^k_{t} \quad \forall j \neq k$}  \Comment{If groups are not balanced}
        \State $n^\star_{t} = n^k_{t}$  \Comment{Set target n. assessments as min.}
        \For {$s \in \{1,...,S\}$}  \Comment{For $S$ Bootstrap simulations}
            \State ${\bm{y}^{j}}^\star_{t,s} \mbox{ } \sim {\bm{y}^{j}}_{t} \quad \mbox{s.t. } \|{\bm{y}^{j}}^\star_{t,s} \| < n^\star_{t}$; \Comment{Resample from majority group }
            \State $\lambda_{t,s}  = \frac{1}{2} \left( \bar{\bm{y}}^{D \star}_{t,s} + \bar{\bm{y}}^{R \star}_{t,s} \right)$  \Comment{Calculate metric for this round}
        \EndFor 
        \State $\nu^{ \{ \mbox{\scriptsize naive}, \mbox{\scriptsize balanced} \} }_{t} = \bar{\lambda}_t$ \Comment{Monte Carlo Mean of simulated metrics}
    \Else
        \State $n^\star_{t} = n^j_{t} = n^k_{t}$ \Comment{Sample is naturally balanced }
        \State $\nu^{ \{ \mbox{\scriptsize naive}, \mbox{\scriptsize balanced} \} }_{t} = \frac{1}{2} \left( \bar{\bm{y}}^D_t + \bar{\bm{y}}^R_t \right)$ \Comment{Calculate metric}
    \EndIf
\EndFor
\end{algorithmic}
\end{algorithm}
\noindent The third naive metric (Equation \ref{eq:eqw_rake}) is a weighted average of the sample, where the weights are derived via raking / iterative proportional fitting (IPF) \citep{deming1940least, kolenikov2014calibrating} to match population marginals. It is referred to as $\nu^{ \{ \mbox{\scriptsize naive}, \mbox{ \scriptsize rake}\} }$. IPF was implemented using the \texttt{anesrake} package from the American National Election Study (ANES) \citep{debell2009computing}. Target marginals are derived from the census and the known $2020$ US Presidential election results. 
Lucid survey quotas enforced representation at the level of age categories and gender -- this left a number of socio-demographic variables, such as $2020$ presidential election vote, unbalanced. Hence raking is necessary to ensure complete conformity with population margins. It is worth noting that the crowd producing assessments for any given tweet, which is small as we discussed earlier, remains unrepresentative under this approach -- it is the full sample of crowd-workers that is weighted using IPF, not the tweet-level assessments. 
\begin{equation}
\label{eq:eqw_rake}
\nu^{ \{ \mbox{\scriptsize naive}, \mbox{ \scriptsize pop.}\} }_{t} = \frac{1}{n_t}  \sum_{i \in \bm{\iota}_t} \omega_i y_{ti} 
\end{equation}
\noindent The final metric we present here (Equation \ref{eq:partisan_accuracy}) relies on crowd-workers' recalled $2020$ vote to identify partisans and produce party-specific scores. The resulting veracity score for each party is referred to as $\nu^{ \{ \mbox{\scriptsize naive}, \mbox{ \scriptsize party}\} }_j$.
These partisan veracity metrics are of interest for two reasons: i. their relationship can indicate cleavages in the underlying understanding of truth -- i.e. a negative correlation in partisan scores indicates Republicans identify as true the same tweets that democrats identify as false, and vice-versa; ii. they function as useful benchmarks for the other metrics, as they reveal the distance of partisan assessments from the other proposed assessment metrics. 

\begin{equation}
\label{eq:partisan_accuracy}
\mbox{Let: } \bm{q}^j_t \subseteq \bm{\iota}_t \quad \forall j \in \{R,D\}; \quad \rightarrow \quad \nu^{ \{ \mbox{\scriptsize naive}, \mbox{ \scriptsize party}\} }_{tj} = \frac{1}{\| \bm{q}^j_t\|}  \sum_{i \in \bm{q}^j_t} y_{ti} 
\end{equation}

\subsection{Model-based Estimation}

Unlike naive approaches, model-based estimation leverages the correlation between: a. socio-demographics characteristics of crowd-workers, the tweets they are assessing, their interactions; and b. the veracity assessment for a given tweet, to learn a model of ``assessment of a given tweet by each rater-persona''. A persona could be, for instance, a Republican male, living in Ohio. The tweet-features used are described below. We can use this model to generate smooth veracity assessments for each tweet under consideration, from the prospective of each persona in the population.

\noindent We fit a Hierarchical Bayesian ordinal logistic regression to model the probability of an individual picking a given veracity score for a given tweet. In the necessarily concise exposition which follows, we borrow from \citep{grilli2014ordered} and \citep{betancourt2019ordinal}. Let $y_{ti}$ be the ordinal, 4-point scale ($C = 4$), veracity assessment described above, for tweet $t$ and crowd-worker $i$. Let $\bm{\pi}_{ti} = \{\pi^1_{ti},...,\pi^C_{ti} \}$ be a vector of probabilities, such that $\pi^c_{ti} = \Pr(y_{ti} = c) \quad \forall c \in \{1,...,C\}$ represents the likelihood that a crowd-worker assesses a tweet with a the $c^{th}$ assessment category. $\psi^c_{ti} = \Pr(y_{ti} \leq c) $ is then a cumulative probability of selecting the $c^{th}$ category. We can express $\pi^c_{ti}$ as a function of cumulative probabilities: $\pi^c_{ti} =\psi^c_{ti} - \psi^{c-1}_{ti} $.
It follows that an ordinal logit model can be expressed as a set of $C-1$ equations (since the cumulative probability of picking the most severe assessment category is  $\psi^C_{ti} = 1$) on the logit scale, as follows:

\begin{align}
\label{eq:ordinal_model}
y_{ti} \sim& \mbox{Categorical}(\pi^1_{ti},...,\pi^C_{ti})\\
\pi^c_{ti} =& \psi^c_{ti} - \psi^{c-1}_{ti};\\
\mbox{logit}(\psi^c_{ti}) =& \alpha_c - \mu_{ti};
\end{align}
where $\alpha_c$ are category-specific threshold values ($\alpha_1 < \alpha_2 <....<\alpha_{C-1}$), and $\mu_{ti}$ is a shared linear predictor.  The global intercept has to be implicitly set to $0$ for identifiability purposes. The linear predictor has negative sign, indicating that positive correlates of veracity decrease the cumulative probability of the `less severe' categorical assessments, and vice-versa increase that of the more severe ones, essentially shifting the whole distribution of veracity closer to the ``\emph{Very Accurate}'' category. The shared linear predictor and the category-wise intercepts together enforce a ``parallel regression'' assumption. The linear predictor is defined as follows:
\begin{align}
\label{eq:net_model}
\mu_{ti} =& \tau^{\scriptsize \mbox{tweet\_id}}_{t} +  \tau^{\scriptsize \mbox{context}}_{z[t]} +  \beta^{gender}_{g[i]} + \beta^{age}_{a[i]}  + \beta^{state}_{s[i]} + \beta^{party}_{p[i]} + \gamma^{context:party}_{k[i]};\\
\theta^{u} \sim \mathcal{N}(0, \sigma^{u}) & \hspace{30pt} \forall \quad \theta \in \{\bm{\tau}, \bm{\beta}, \gamma\}; \quad u \in \{\text{tweet, context, gender, \ldots} \mid \exists\ \theta^u\};\\  \sigma^{u} \sim \mathcal{N}^{+}(0,1).
\end{align}
\noindent $\mu_{ti}$ incorporates information from tweets and crowd-workers. Worker characteristics such as $gender$, $age$, $state$, and $party$, are given partially pooled priors to leverage shrinkage in and avoid over-fitting on the self-selected sample of Lucid respondents. These effects capture workers' inherent likelihood to rate any post as accurate or not -- e.g. young participants are more skeptical about the veracity of any given tweet than older participants. $context$ effects refer to the impact of context annotation assigned to tweets as explained in Section \ref{sc:external data}. The context parameter is included to capture any inherent inclination of the sample towards a given topic. It is interacted with the $party$ effect to capture special predispositions of a political group within a given context -- e.g. the average veracity score of Republicans on tweets mentioning the White House may differ from the Democrats' on the same context-annotation, and may further be different from the average Republican veracity score for tweets mentioning Ron DeSantis. Although the $context:party$ interaction could be extended to all other covariates (e.g. age-context interaction),  this would drastically increase the required statistical and computational power. The $context:party$ interaction was chosen due to the highly political nature of fake news and the selected tweets. The remaining $tweet\_id$ parameter then represents the residual tweet veracity, ceteris paribus.\\

\noindent Similarly to naive veracity metrics, $sample$, $balanced$, $population$ and $partisan$ veracity scores were calculated for all tweets. For computational reasons we collapse the uncertainty of the Bayesian posterior predictions into point-estimates (typically the Monte Carlo Mean of the simulated posterior), though in principle one could recover the full posterior of each of these metric. 


   


Model-based sample veracity scores are calculated by taking the simple average of fitted values:
\begin{equation}
\nu^{ \{ \mbox{\scriptsize model}, \mbox{ \scriptsize sample}\} }_{t} = \frac{1}{n_t} \sum_{i \in \bm{\iota}_t} \hat{y}_{ti}.
\end{equation}
\noindent Model-based balanced veracity scores necessitate the computation of model-based partisan veracity scores, hence we present both below. The partisan scores are computed via a Multilevel Regression and Post-stratification (MrP, \cite{park2004bayesian}) approach, as follows: i. we predict a veracity score, for each tweet, for each Republican or Democrat persona in the population; ii. take the weighted average of these predicted values where the weights are defined by the number of individuals from each `persona'-type in the population. This metric is presented in Equation \ref{eq:eqw_model_party}: 
\begin{equation}
\label{eq:eqw_model_party}
   \nu^{ \{ \mbox{\scriptsize model}, \mbox{ \scriptsize party}\} }_{tj} =\frac{\sum_{h \in \mathcal{P}^j} \omega_h \hat{y}_{th} }{\sum_{h \in \mathcal{P}^j} \omega_h }, \quad \forall j \in \{R,D\};
\end{equation}
where $\mathcal{P}^j$ represents the space of all possible personae in the population, with each persona indexed by $h$. The superscript $j$ indicates we exclusively look at personaes who voted for party $j$ in $2020$. We can then use the partisan estimates to generate a balanced score; this is detailed in Equation \ref{eq:eqw_model_balanced}:
\begin{equation}
\label{eq:eqw_model_balanced}
\nu^{ \{ \mbox{\scriptsize model}, \mbox{ \scriptsize balanced}\} }_{t} = \frac{1}{2} \left( \nu^{ \{ \mbox{\scriptsize model}, \mbox{ \scriptsize party}\} }_{t, j = R} +\nu^{ \{ \mbox{\scriptsize model}, \mbox{ \scriptsize party}\} }_{t,j=D}  \right).
\end{equation}
\noindent The model-based population veracity (Equation \ref{eq:eqw_model_pop}) is computed similarly to the partisan scores, in that it necessitates post-stratification. The difference is that instead of fitting the model to the partisan personae in $\mathcal{P}^j$, we fit to every persona in the population $\chi$, where $\mathcal{P}^j \subseteq \chi$, and take the weighted average accordingly.  
\begin{equation}
\label{eq:eqw_model_pop}
   \nu^{ \{ \mbox{\scriptsize model}, \mbox{ \scriptsize pop.}\} }_{t} =\frac{\sum_{h \in \chi} \omega_h \hat{y}_{th} }{\sum_{h \in \chi} \omega_h };
\end{equation}
Given the weights represent the prevalence of each persona in the population, this is an estimate which can be understood as a  synthetic, democratic, ``one man, one vote'' style decision on the veracity of the tweet. 

\noindent The value of model-based metrics is two-fold: i. they reduce overfitting to the small, unrepresentative sample of veracity assessments for each tweet; ii. (with the exception of $\nu^{ \{ \mbox{\scriptsize model}, \mbox{ \scriptsize sample}\} }_{tj}$) they aggregate the perspectives of all personae within the stratification frame. Unlike naive metrics, which depend solely on a limited number of reviews collected through surveys per tweet, model metrics utilize the predicted values across the entire population, ensures a more comprehensive representation of diverse viewpoints, enhancing the reliability and robustness of the veracity scoring process.

\subsubsection{Model Fitting}
The model was implemented via the probabilistic programming language Stan \citep{gelman2015stan,carpenter2017stan}. It utilizes a binary tree depth of $10$ and was run with a total of $7$ chains worth $500$ post-warm-up iterations each. 
All chains and parameters of the model converged ($\hat{R} \le 1.05$, $N_{eff}/N \ge 10\%$, $N_{eff} \ge 400$ , $se_{mean}/\sigma{} < 10\%$). The computed model, and visualizations of the posterior distributions of the parameters, can be found in the supplementary material, the necessary code to run the model is also provided but requires several days of computation on a personal computer\footnote{Intel i5 with 8GB of RAM}.

\subsubsection{Model Interpretation}
Table 1 in the Appendix presents the results of the ordinal logit model. There is no evidence for general socio-demographic effects on the propensity to label a tweet as true or accurate. Credible intervals for all socio-demographic main-effects are consistently centred around $0$. 

\begin{figure}[htp]
   \centering
    \includegraphics [scale = .7] {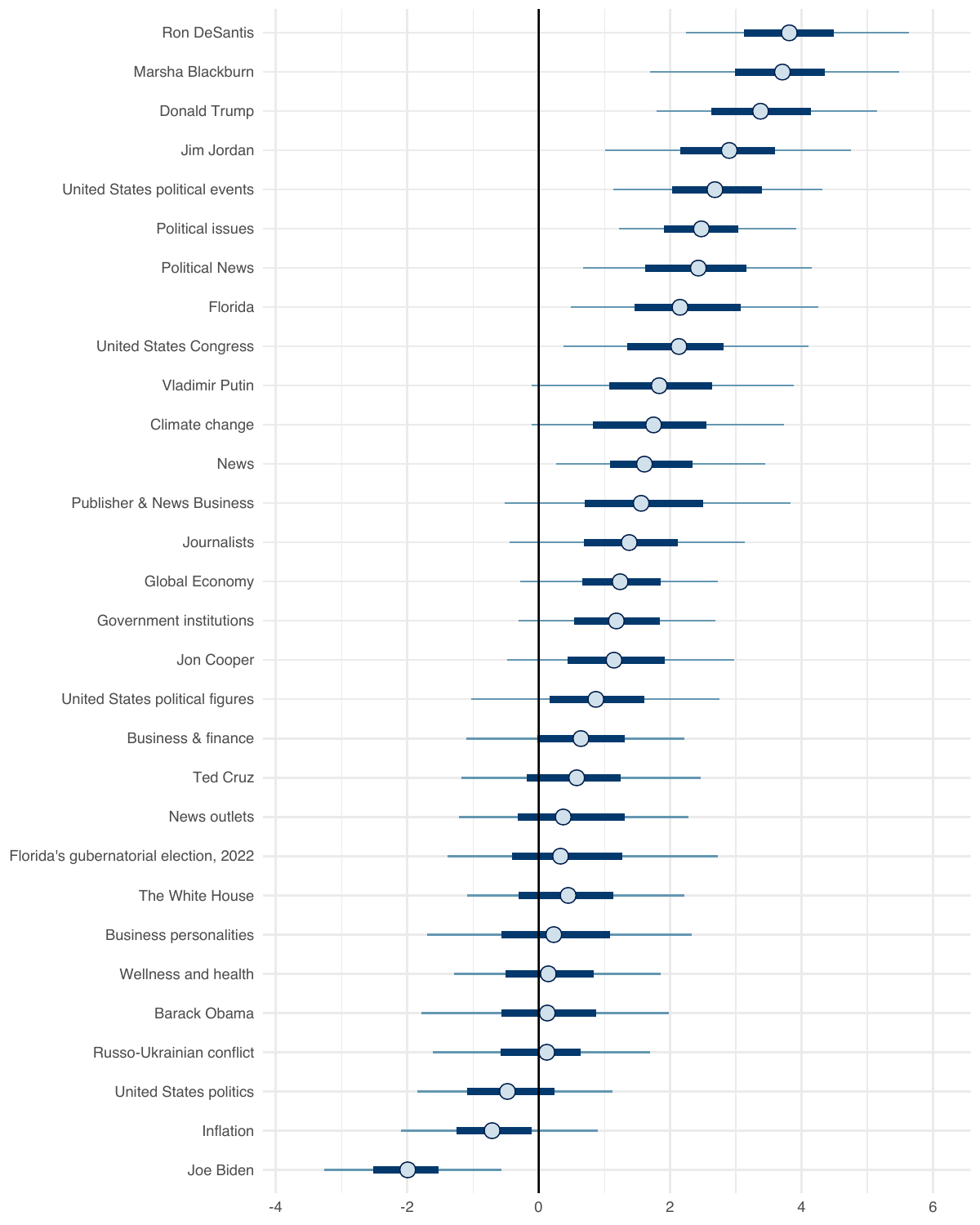}
      \caption{Difference (in logit-scale predicted values) between Democrat and Republican predispositions towards the selected context annotations. Negative/Positive values respectively indicate that Republicans/Democrats are more likely to rate the topic as `true'. Note that most tweets tend to have negative sentiment, hence the partisan divide is often driven by believing negative facts about a political opponent are true.} 
   \label{fig:conint}
\end{figure}

Notably, interaction effects between political affiliation and context were statistically significant, indicating partisans differed in their understanding of ``truth'' across context annotation. This suggests that both the overall crowd and its political subgroups displayed predispositions towards specific topics
Figure \ref{fig:conint} visually illustrates the difference in the logged odds of Democrats and Republicans for the chosen annotations.

\subsection{Comparing Veracity Metrics}
\begin{figure}
    \centering
    \includegraphics [scale =0.55] {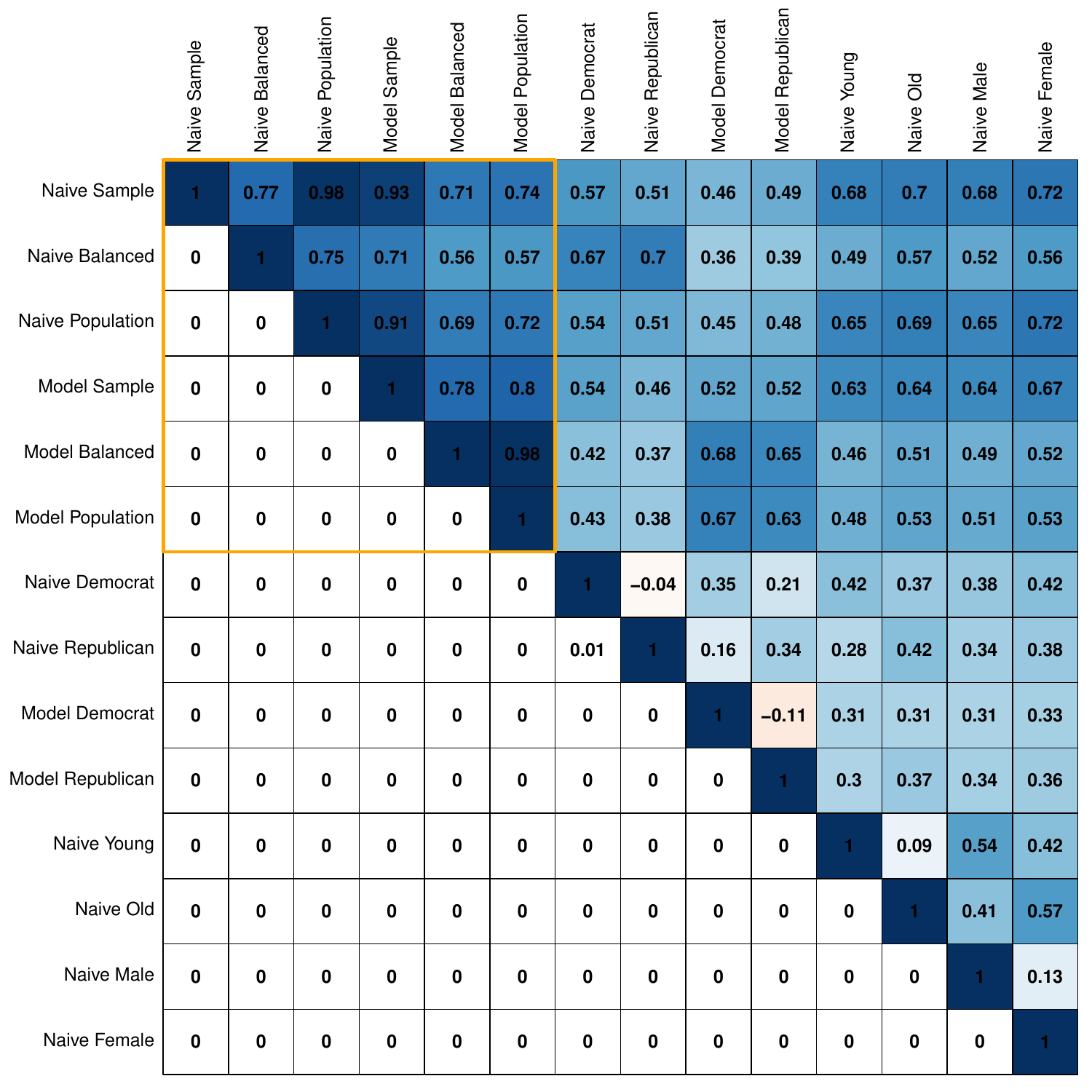}
    \caption{The upper half of the matrix indicates the correlation coefficients; the lower half the p-values of the respective correlations. The yellow rectangle highlights the wisdom of the crowd metrics, non-highlighted metrics are unrepresentative of the crowd by design.}
    \label{fig:cormat}
\end{figure}

The correlations across all accuracy estimates are summarized in Figure \ref{fig:cormat}. Estimating the correlations across all metrics serves multiple purposes. First, it validates the model-based estimates. The high correlation between $\nu^{ \{ \mbox{\scriptsize naive}, \mbox{ \scriptsize sample}\} }$  and $\nu^{ \{ \mbox{\scriptsize model}, \mbox{ \scriptsize sample}\} }$ indicates that the model can replicate the assessments made in the survey to a significant extent.
Second, the correlations offer an indication of the consistency and agreement between different evaluation approaches. If two metrics are highly correlated, it suggests that they tend to yield similar results. 
All metrics exhibited significant correlations except for the partisan metrics. All naive and model-based correlations were significant, indicating consistency across both methods. 

The partisan metrics stand-out as the only negatively-correlated veracity scores amongst all that we evaluate. This is a chilling finding -- it exposes clearly the underlying polarization around fake news sharing, and notions of truth, online. 
Table \ref{tab:extweets} displays $4$ example tweets along with their accuracy scores, it exemplifies the differences and similarities in Democrat and Republican ratings along with the inner workings of naive and model-based scores. In contrast to the partisan metrics, analyses of other dichotomous classifications (i.e., young and older reviewers, or female and male reviewers) reveal positive and statistically significant correlations across respective categories. We can conclude that partisanship plays a key, definitional role in the understanding of fake news. 
This lack of consensus across political groups 
threatens efforts to endow fake news detection with democratic legitimacy.


\begin{table}[htp]
    \centering
    \begin{tabular}{p{9cm}p{5cm}}
        \toprule
        \textbf{Tweet Text} & \textbf{Accuracy Metrics} \\
        \midrule
        Vaccines have been saving us for centuries so why stop being vaccinated now?! we took vaccines for chickenpox measles polio! Without these vaccines millions would have died! What’s any different than this from Covid!? And those that don’t take it make others vulnerable! &
        \begin{tabular}[t]{@{}lr@{}}
            \textbf{Naive Sample} & 3.21 \\
            \textbf{Model Population} & 3.72 \\
            \textbf{Only Democrats} & 4.00 \\
            \textbf{Only Republicans} & 2.57 \\
        \end{tabular} \\
        \midrule
        That’s for 2022, general gas prices Pre COVID where around \$2 to \$2.50 that’s an increase of 61\% when looking at 3.71. Get out inflation is killing us and they keep adding more dollars with no productivity back up in the economy. &
        \begin{tabular}[t]{@{}lr@{}}
            \textbf{Naive Sample} & 3.25 \\
            \textbf{Model Population} & 3.60 \\
            \textbf{Only Democrats} & 4.00 \\
            \textbf{Only Republicans} & 4.00 \\
        \end{tabular} \\
        \midrule
        In case anyone has forgotten, we have rampant crime, 9\% inflation, blown-up retirement savings, an open border with millions pouring across it illegally, soaring interest rates, COVID-19 mandates and government assaults on moral decency and biology. No joke, man. &
        \begin{tabular}[t]{@{}lr@{}}
            \textbf{Naive Sample} & 2.82 \\
            \textbf{Model Population} & 3.46 \\
            \textbf{Only Democrats} & 1.67 \\
            \textbf{Only Republicans} & 4.00 \\
        \end{tabular} \\
        \midrule
        The state of Florida needs experience and steady hands to lead the recovery after the devastating hurricane and the Covid 19 pandemic. So many lives have been lost because of an inept governor in Ron Desantis and the destruction is unspeakable. We need Charlie Crist and Val Demm. &
        \begin{tabular}[t]{@{}lr@{}}
            \textbf{Naive Sample} & 2.30 \\
            \textbf{Model Population} & 2.10 \\
            \textbf{Only Democrats} & 3.33 \\
            \textbf{Only Republicans} & 1.33 \\
        \end{tabular} \\
        \bottomrule
    \end{tabular}
    \caption{Mentions (e.g. @WhiteHouse) were removed from the tweet for readability.}
    \label{tab:extweets}
\end{table}

\pagebreak

\section{Who is Likely to Spread Fake News ?}\label{map}
We employ the wisdom of the crowd ratings, as calculated in Section \ref{sc:accuracy}, as dependent variables to study the characteristics of individuals who share fake news, and identify areas most at-risk, in the United States. To do so we leverage Multilevel Regression with Post-stratification (MrP). The adoption of MrP is motivated by three considerations: i. the need to address disparities between online and real-life samples \citep{wang2015forecasting,cerina2021polling,cerina2020measuring} in generating area-level estimates; ii. the need to generate accurate area-level projections of the dependent variable of interest \citep{park2004bayesian,lauderdale2020model}; iii. the necessity to interpret the fitted model of fake-news sharing, by studying estimated coefficients across individual and area levels. 
Though commonly used for pre-election polling, MrP has been shown to be appropriate for any number of socio-scientific outcomes of interest e.g. to predict area-level anti-migrant sentiment \citep{butz2016estimating}, support for gay marriage \citep{gelman2016using} and the decline of religiosity over time \citep{wiertz2021rise}. In the context of social media, MrP offers a means to account for the compositional differences between online and real-life environments \citep{mellon2017twitter, mislove2011understanding}.

\noindent To be explicit, the use of MrP implies we are generalising the behaviour of fake news sharing on $\mathbb{X}$ across the whole population, including across those strata which are not on $\mathbb{X}$. We assume implicitly that if someone shares a fake piece of digital content on $\mathbb{X}$, they're equally likely to do so offline, and vice-versa, ceteris-paribus. This assumption should be empirically tested -- we leave this for future work.

\noindent Whilst we cannot fully account for online selection via MrP, we can account for selection on socio-demographic variables. Specifically, online samples are found to be younger, more educated and liberal-leaning \citep{mellon2017twitter} -- we can account for these demographic imbalances via MrP -- i.e. by accounting for these correlates of fake news sharing in our regression model, and post-stratifying predictions to a representative stratification frame.  


\subsection{Modelling the Propensity to Share Fake News}

Resetting our notation convention for this part of the paper, let $i \in \{1,..,n\}$ be an index of $\mathbb{X}$ users who shared at least one tweet in our sample of tweets. The goal is to produce state-level estimates of the propensity to share fake-news, which are directly interpretable as the proportion of people in the state who are at risk of sharing fake news. Hence we consider, for each user, the set of tweets they have shared $\tau_i  \in \{\tau_1,...,\tau_n\}$. Let $\bm{\nu}^{m}_{\tau[i],i}$ be a vector of veracity scores belonging to tweets shared by individual $i$, for a given metric $m$. This is our dependent variable -- the analysis which follows is repeated for all of the veracity metrics described in previous sections.

\subsubsection{Dichotomisation}
In the interest of clarity, we transform our dependent variables to be immediately interpretable as sharing `fake news', rather than an ordinal assessment of veracity. We collapse our veracity metrics into dichotomous scores $\bm{\nu}^{m,\star} \in \{0,1\}$, where a value of $1$ indicates an instance of fake news; the dichotomising threshold is chosen to be $\nu\leq 2$, or in words ``\emph{Not very accurate or less}''

\[
\nu^{\star} =
\begin{cases} 
1 & \text{if } \nu \leq 2 ;\\
0 & \text{otherwise}.
\end{cases}
\]
\noindent The choice of ``\emph{Not very accurate}'' seems a natural threshold, in that it represents the first item (starting from ``\emph{Very Accurate}'') on the likert scale indicating some level of negative judgment on the accuracy of the underlying tweet. Beyond natural interpretation, fixing a definition of truth exogenously enables making cross-model comparisons in the analysis which follows. The most consequential drawback of this choice is its sensitivity to extreme assessments\footnote{There are other concerns: i. metrics which require exclusion of assessments, for example partisan or balanced metrics, will have an extraordinarily low-$n$ per tweet (e.g. $\sim 3$ assessments), leading to noisy veracity scores; ii. for model-based veracity scores, due to a ``regression to the mean'' effect, we observe scores which are closer to the center of the likert scale, leading to an attenuation bias effect that somewhat artificially reduces the number of fake news.}. Due to the veracity score being an average of the underlying veracity assessments, outlier assessments can have a disproportionate effect on the chances of a given tweet being labeled as fake. In the context of social media content moderation, and the anti-correlation between partisans' understanding of truth, this carries political consequences: if a partisan group appears more likely to flag any tweet as ``fake'', whilst the opposing partisans are more ambivalent in their assessments, the ultimate judgment will tend to favour the side which is most harsh about the degree of (in)accuracy of the tweet. This could therefore bias our analysis of the kinds of users who share fake news by favouring the account of the partisans described above.

To address this concern, we check the robustness of our inference under a second dichotomisation strategy -- namely one which takes ``the least accurate $10\%$'' of tweets to be fake, under each metric. This then becomes a relative measure, quasi-fixing the baseline sharing propensity $\sim 10\%$ by definition, and balancing out the ``intensity'' of partisan assessments, at the cost of losing the ability to compare fake news sharing propensity across models. From the prospective of the social media company implementing a fake news detection tool, this is an attractive strategy: they can retain the claim that users of `every kind' (all the personae in the population) were included in the assessment, whilst addressing the issue of different ``intensity'' of fake news detection across partisans. The strategy is akin to implement a system of ``tokens'', where partisans get $10$ ``fake news'' tokens to distribute across every $100$ tweets they review.



\subsubsection{Model Specification}
\noindent Given our now dichotomous dependent variable, we fit a Hierarchical Bayesian logistic regression model to the data, with the intention of estimating the parameters which determine the propensity to share fake news. The independent variables of the multilevel model featured both user- and state-level information. User gender, age and party-id were gathered through the use of external tools as outlined in section \ref{sc:external data}. The states of residency for each user were inferred through the geo-tags provided by Twitter.

We create a state-level predictor with a relatively small number of theoretically motivated covariates. We include the percentage of people who identify as being of white ethnicity -- this is a catch-all proxy for a number of relevant correlates, such as rurality, diversity within the state relative to the majority group, and ideological conservatism of the state. The inclusion of this variable is motivated largely by the context of the tweets we have in our corpus -- we expect that as states get `whiter', less diverse, more rural and more conservative, people will tend to be more at risk for believing conspiratorial narratives about the pandemic, as opposition and skepticism towards federal pandemic policies was strong within these states. We further include the \% of people within the state who own a college degree -- this is motivated by the idea that a well educated polity is endowed with a sort of ``herd immunity'' towards misinformation, and we would therefore expect less fake news sharing in more highly educated states. Finally, we include population density in the model. This measure is in our view a good correlate to the average size and diversity of the networks of individuals living in the state. We expect people who live in denser states to be exposed to a variety of narratives and perspectives, which may endow them with a greater ability to discern between true and fake information, and act as a risk-reduction factor to the spread of fake news.

The percentage of white residents and the proportion of those with a college degree  were derived from census data \citep{ipumsusa}. The stratification frame uses data from IPUMS USA \citep{ipumsusa} and the American National Election Studies (ANES) \citep{anes}. Since the joint distributions of party-id and socio-demographics are unavailable from the census, the stratification frame was extended with the aid of a synthetic joint distribution (MrsP) \citep{leemann2017extending}. The full model and the post-stratification procedure are detailed in Equations \ref{eq:step2_model_1} to \ref{eq:step2_model_2}. 

\begin{align}
\bm{\nu}^\star_{\tau[i],i} \sim & \mbox{Bernouilli}(\theta_{i});\label{eq:step2_model_1} \\ 
\mbox{logit}(\theta_i) = & \alpha + \\
& \beta^{gender}_{g[i]} + \beta^{age}_{a[i]} + \beta^{party}_{p[i]}  + \beta^{state}_{s[i]} + \\
& \gamma_1 \mbox{white\%}_{s[i]}  +\gamma_2 \mbox{college\%}_{s[i]} + \gamma_3 \mbox{pop.density}_{s[i]};\\
\alpha \sim & \mathcal{N}(0,1); \\
\beta^{u} \sim&  \mathcal{N}(0, \sigma^{u}), \hspace{30pt}\forall \hspace{2pt} u \in \{\text{gender, age, party, state}\};\\
\sigma^{u} \sim & \mathcal{N}^{+}(0,1);\\
\bm{\gamma} \sim &  \mathcal{N}(0,1); \\
\hat{\theta}_h = \mbox{logit}^{-1}( &\hat{\alpha} + \\
& \hat{\beta}^{gender}_{g[h]} + \hat{\beta}^{age}_{a[h]} + \hat{\beta}^{party}_{p[h]}  + \hat{\beta}^{state}_{s[h]} + \\
& \hat{\gamma}_1 \mbox{white\%}_{s[h]}  + \hat{\gamma}_2 \mbox{college\%}_{s[h]} + \hat{\gamma}_3 \mbox{pop.density}_{s[h]}) ;\\
\hat{\theta}^\star_{s} = & \frac{\sum_{h \in \chi} \omega_h \hat{\theta}_h}{\sum_{h \in \chi} \omega_h}.
\label{eq:step2_model_2}
\end{align}

\noindent Where $h$ is a cell (or a stratum, or persona) in the stratification frame, $s$ is a state identifier, and $\hat{\theta}_{h}$ is the predicted probability of sharing fake news for cell $h$. $\hat{\theta}^\star_{s} $ represents the expected proportion of individuals in state $s$ who are at risk of sharing fake news. The model is again implemented via Stan \citep{gelman2015stan,carpenter2017stan}. A total of $10$ chains with $4000$ post-warm-up iterations were run with a binary tree depth of $15$ \footnote{$\nu^{\{ \mbox{\scriptsize naive}, \mbox{ \scriptsize Republican}\}}_{j=R}$ and $\nu^{\{ \mbox{\scriptsize naive}, \mbox{ \scriptsize Balanced}\}}$ used a binary tree depth of 20 to ensure the convergence of the model parameters.}. 
A summary of model coefficients fit to each veracity metric is provided in the supplementary materials, and their fully computed form is available online on \texttt{GitHub}. All models fully converged with chains mixing and low autocorrelation ($\hat{R} \le 1.05$, $N_{eff}/N \ge 10\%$, $N_{eff} \ge 400$, $se_{mean}/\sigma{} \le 10\%$). 

\subsection{Fake News Sharing Inference}
Models fit across veracity metrics 
generally yield similar coefficient estimates. The findings presented below are generally robust irrespective of dichomotisation strategy. Coefficients estimated using model-based veracity scores as dependent variables tend to be somewhat attenuated relative to those estimated with naive veracity scores. This is to be expected as model-based veracity scores experience a ``regression to the mean'' effect which affects the overall variance of these metrics. Unless otherwise specified, effects in the following paragraphs are presented for models fit to the model-based population metric -- or the ``wisdom of the synthetic \& representative crowd'' under the $\nu\leq2$ dichotomisation strategy. Effect-sizes are communicated via $80\%$ credibility intervals. Detailed tables including summary statistics for every model fit to each veracity metric are available in the Appendix. 

The probability of sharing fake news is generally quite low across models -- fitting the model to $\nu^{\{\mbox{\scriptsize model}, \mbox{\scriptsize pop.}\}}$ yields a baseline probability of sharing fake news of $[0.07,0.14]$, which is generally similar across metrics. A notable exception appears when fitting to $\nu^{\{\mbox{\scriptsize model}, \mbox{\scriptsize pop.}\}}_{j=D}$ exhibited somewhat higher baseline propensity ($[0.12,0.29]$), making it roughly twice as likely an individual would be labeled at-risk of sharing fake news if Democratic partisan definition of fake news are used, net of all other characteristics. Using $\nu^{\{\mbox{\scriptsize model}, \mbox{\scriptsize balanced}\}}$ presents an interesting effect -- namely, balancing partisan assessments leads to a substantial attenuation in the estimated \% of individuals sharing fake news, as shown by a lower baseline propensity $[0.04,0.13]$.

\begin{figure}[htp]
    \centering
    \begin{subfigure}{0.44\textwidth}
        \centering
        \includegraphics[width=1\linewidth]{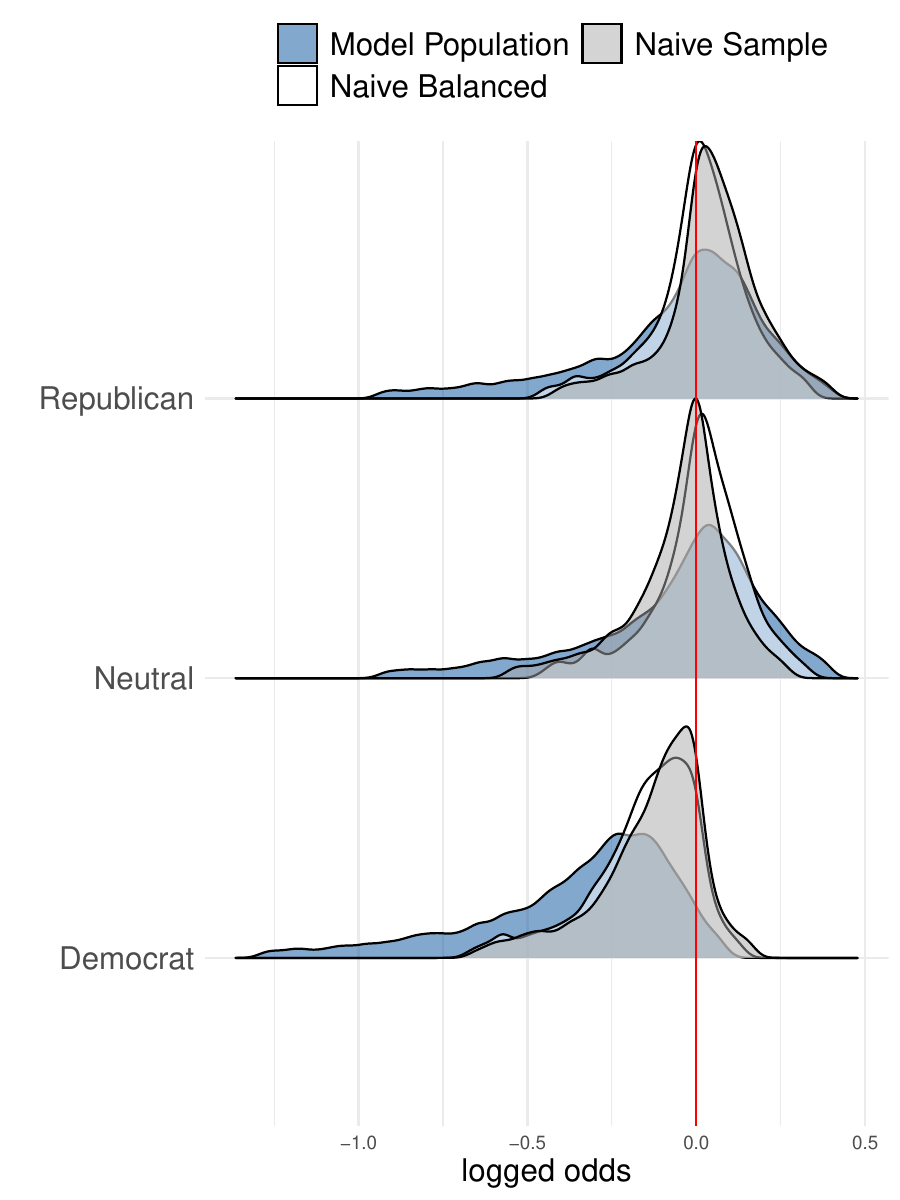}
        \caption{Wisdom of the Crowds Metrics}
        \label{fig:step2_politics}
    \end{subfigure}
    \hfill
    \begin{subfigure}{0.44\textwidth}
        \centering
        \includegraphics[width=1\linewidth]{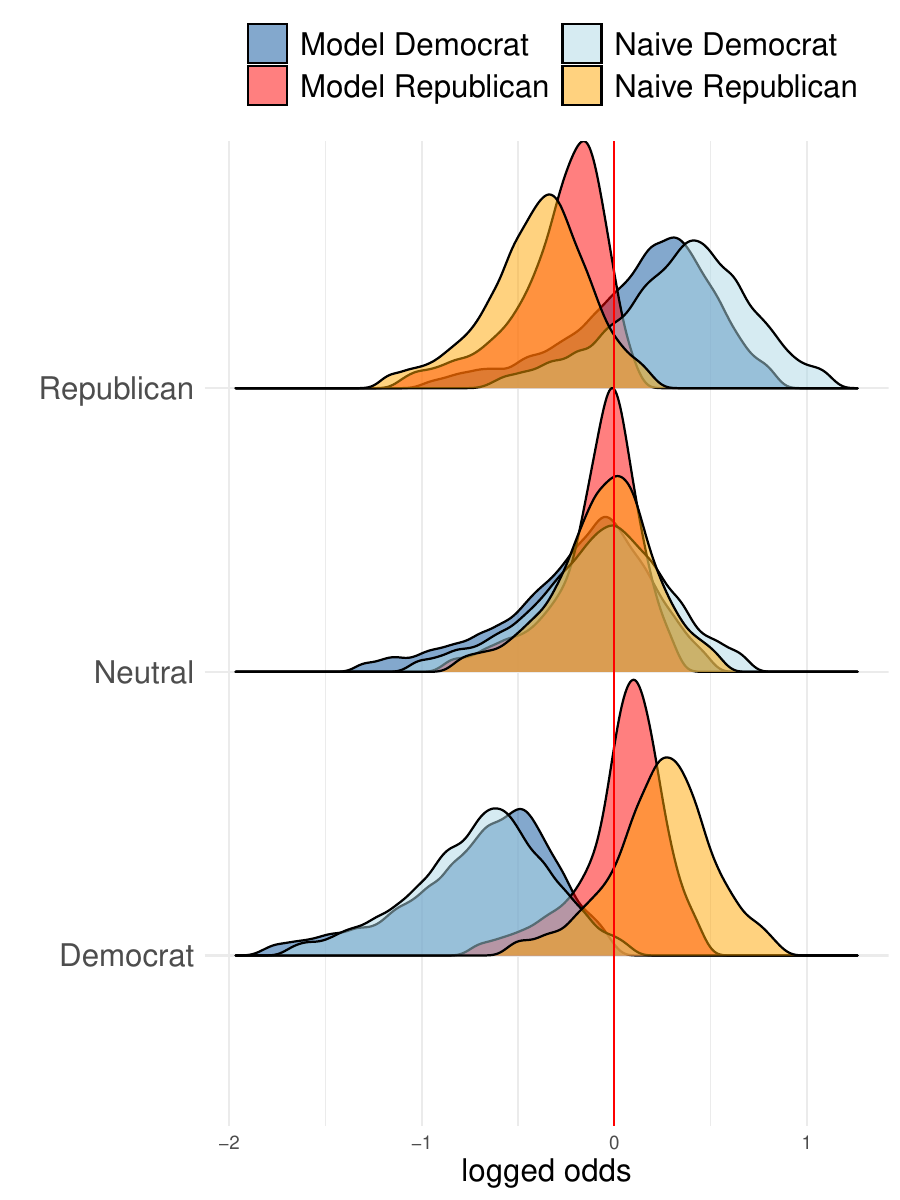}
        \caption{Democrats and Republicans}
        \label{fig:step2_repdem}
    \end{subfigure}
    \caption{Posterior distribution of the party random-effects on fake news sharing estimation. Negative values indicate a reduced likelihood of posting fake news. The legends identify the veracity score used to estimate a given effect.}
\end{figure}

The most striking finding from this analysis is around the role of partisanship. The posterior distributions of the partisanship random effects are presented in Figure \ref{fig:step2_politics}.  
Across metrics, Democrats consistently demonstrated a lower likelihood of sharing fake news on $\mathbb{X}$, relative to the average user. Using $\nu^{\{\mbox{\scriptsize model}, \mbox{\scriptsize pop.}\}}$ as the dependent variable, being a Democrat is associated with a reduction in the odds of spreading fake news of anywhere in $[57.3\%,3.9\%]$. This finding is robust in that even when fake news are defined giving both Democrat and Republican assessments equal weights, as in $\nu^{\{\mbox{\scriptsize model}, \mbox{\scriptsize balanced}\}}$, being a Democrat is still a strong risk-reduction factor -- $[73.8\%,7.7\%]$. Strikingly, this relationship reverses when Republicans define fake news, as is the case with $\nu^{\{\mbox{\scriptsize model}, \mbox{\scriptsize party}\}}_{j = R}$. Here the Democratic partisan effect on sharing becomes insignificant, whilst Republicans gain a protective effect against sharing fake news of roughly equal size $[50.8\%,2.0\%]$, in terms of odds-reduction. This reversal is a direct consequence of the negative correlation between partisan veracity metrics.

Though partisanship produces the strongest individual-level effects in the model, we find some evidence that sex plays a role in fake news sharing propensity. Being a woman was associated with a reduction in the odds of sharing fake news of $[29.5\%,4.9\%]$. This too was robust to partisan metrics, holding statistical significance and direction in most model-based partisan and balanced metrics.  

As a final note on the individual-level determinants of fake news sharing, we find some relatively weak evidence to the effect that individuals who are older than $40$ years of age are at higher risk of spreading fake news. Here the Finding is more mixed -- in our reference model, that trained on $\nu^{\{\mbox{\scriptsize model}, \mbox{\scriptsize population}\}}$, this effect appears insignificant; however on models trained on $\nu^{\{\mbox{\scriptsize model}, \mbox{\scriptsize party}\}_{j=R}}$, and also in $\nu^{\{\mbox{\scriptsize model}, \mbox{\scriptsize balanced}\}}$ and $\nu^{\{\mbox{\scriptsize model}, \mbox{\scriptsize population}\}}$ when the threshold rule is set to the lowest $10^{th}$ percentile, we find being old is a significant risk factor. Taking as reference $\nu^{\{\mbox{\scriptsize model}, \mbox{\scriptsize population}\}}$, determined using a $10\%$ threshold instead of the usual $\nu\leq2$, we find being above $40$ years of age increases the odds of sharing fake news by $[-5.0\%,31.0\%]$ -- the negative lower-end of the estimate implies a substantial degree of uncertainty. 

\noindent At the state-level, we don't detect much in the way of statistically meaningful effects for our state-level predictors. The only consistently significant correlate is the \%  of white people in a state -- living in a state which is one standard deviation ``whiter'' than the average state is associated with an increase in the odds of sharing fake news of $[-1.9\%,17.4\%]$. The other effects are largely insignificant, though their point estimates are generally in the expected directions across models. 

\noindent In terms of log-odds, the scale of the area-level unexplained heterogeneity ($\sigma_{State}$) for the model fit to $\nu^{\{\mbox{\scriptsize model}, \mbox{\scriptsize population}\}}$ has 80\% interval $[0.06,0.31]$, making it a relatively small source of variation. For comparison, the largely insignificant variation at the level of age categories has scale  $\sigma_{Age}$ in $[0.02,0.29]$, whilst the consequential variation between partisan categories  $\sigma_{Party}$ lies in $[0.13,0.93]$.

\subsection{State-level Share-Risk Estimates}
\begin{figure}
    \centering
    \includegraphics[scale =0.6]{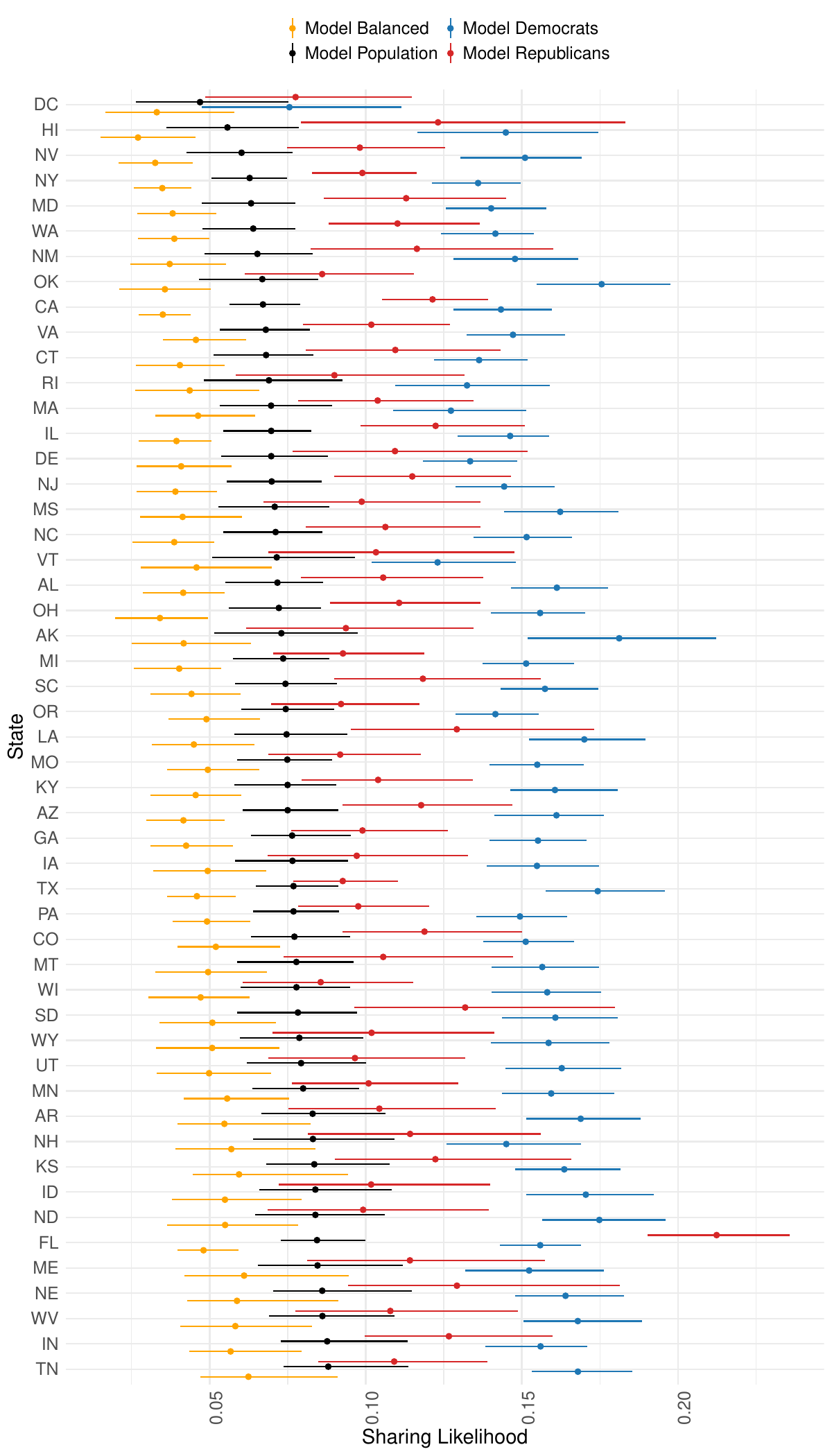}
    \caption{Posterior distribution of state-level \% of individuals who share fake news. }
    \label{fig:state_post}
\end{figure}

Our persona-level predictions are driven primarily by partisanship, and to some degree by the sex of the $\mathbb{X}$ user. Sex is roughly equally distributed across states, so its effect on state-level spread-risk estimates will be minimal. The \% white population in the users' state of residency contributes to some state-level heterogeneity. We post-stratify these predictions into state-level estimates of the \% of fake news sharing. Figure \ref{fig:state_post} presents the posterior distribution of the state-level share of the population who shares fake news. 

\noindent Variance across states is limited, with point estimates ranging $[5\%,8\%]$ for estimates derived from the model-based population veracity score, and $[3\%,6\%]$ for estimates derived from the model-based balanced veracity score. Point estimates of state-level share of fake news sharing population relying on partisan veracity metrics present larger heterogeneity: $[7\%,23\%]$ for estimates derived from the model-based Republican veracity score, $[7\%,18\%]$ for its Democrat counterpart. Note that intervals for partisan estimates are substantially larger, and point estimates more disperse, owing in part to noise due to the smaller pool of tweets available for these models. 

\begin{figure}
    \centering
    \begin{subfigure}{0.45\textwidth}
        \centering
        \includegraphics[width=\linewidth]{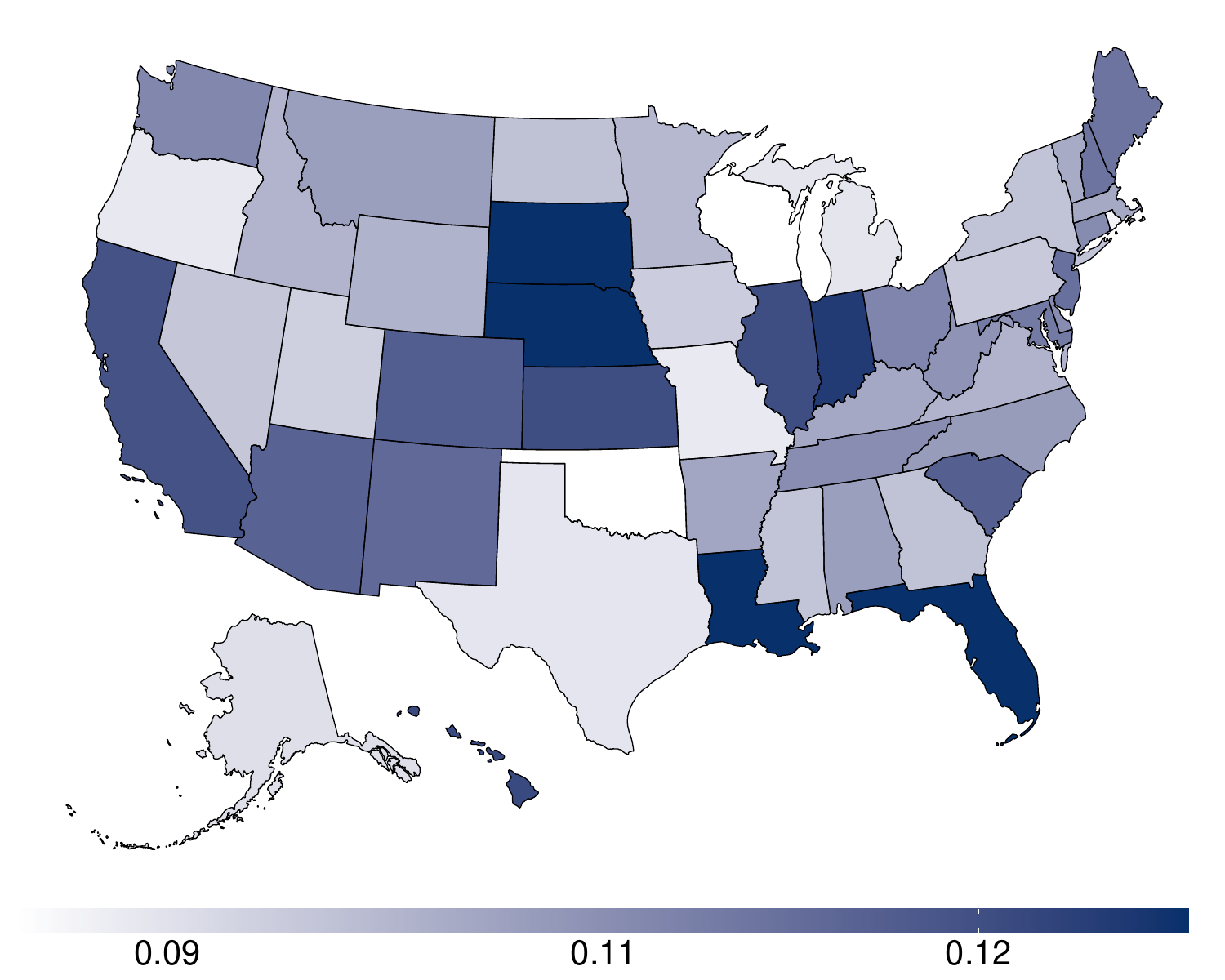}
        \caption{Model Republicans}
        \label{fig:repUSA}
    \end{subfigure}
    \hfill
    \begin{subfigure}{0.45\textwidth}
        \centering
        \includegraphics[width=\linewidth]{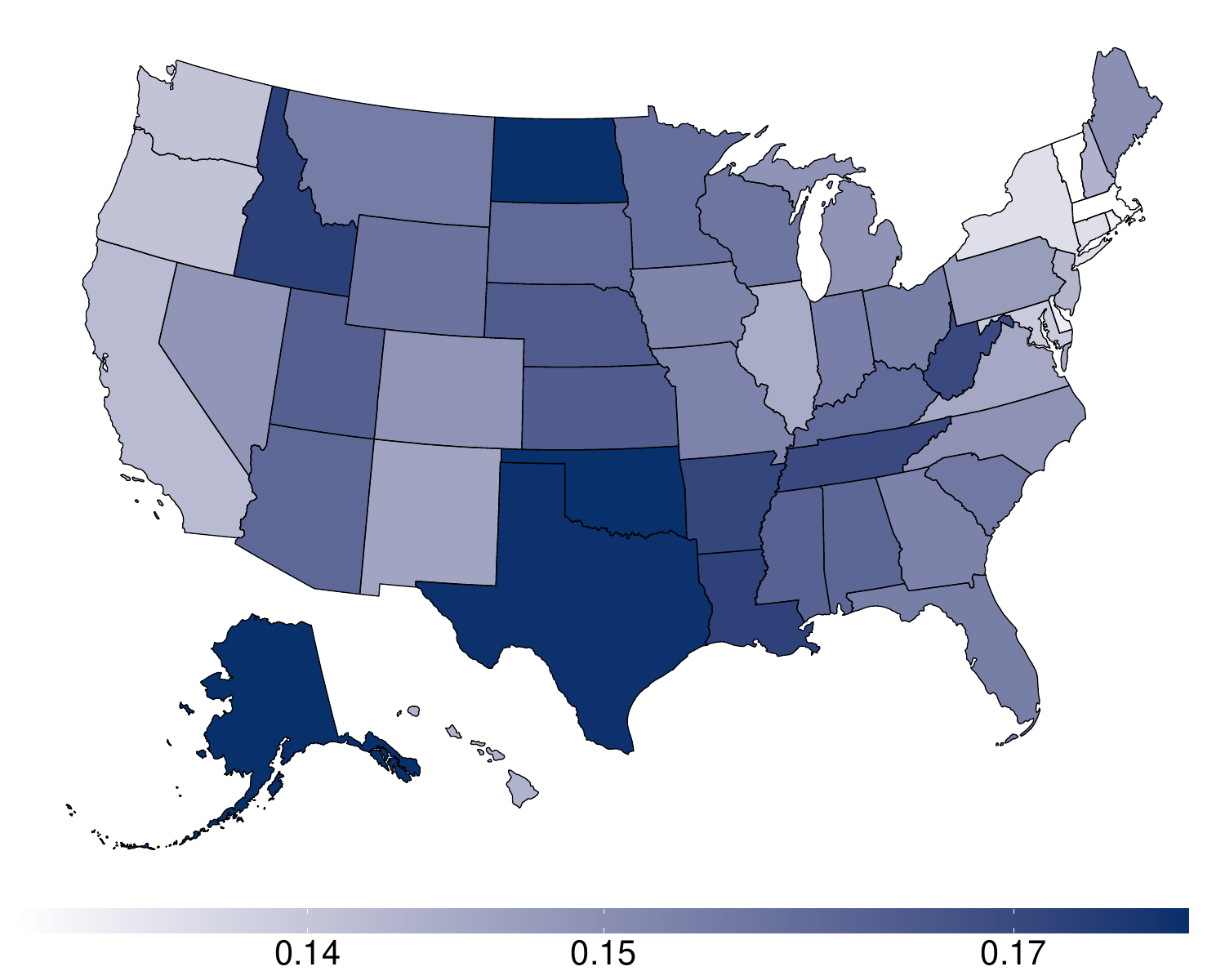}
        \caption{Model Democrats}
        \label{fig:demUSA}
    \end{subfigure}
     \hfill
    \begin{subfigure}{0.45\textwidth}
        \centering
        \includegraphics[width=\linewidth]{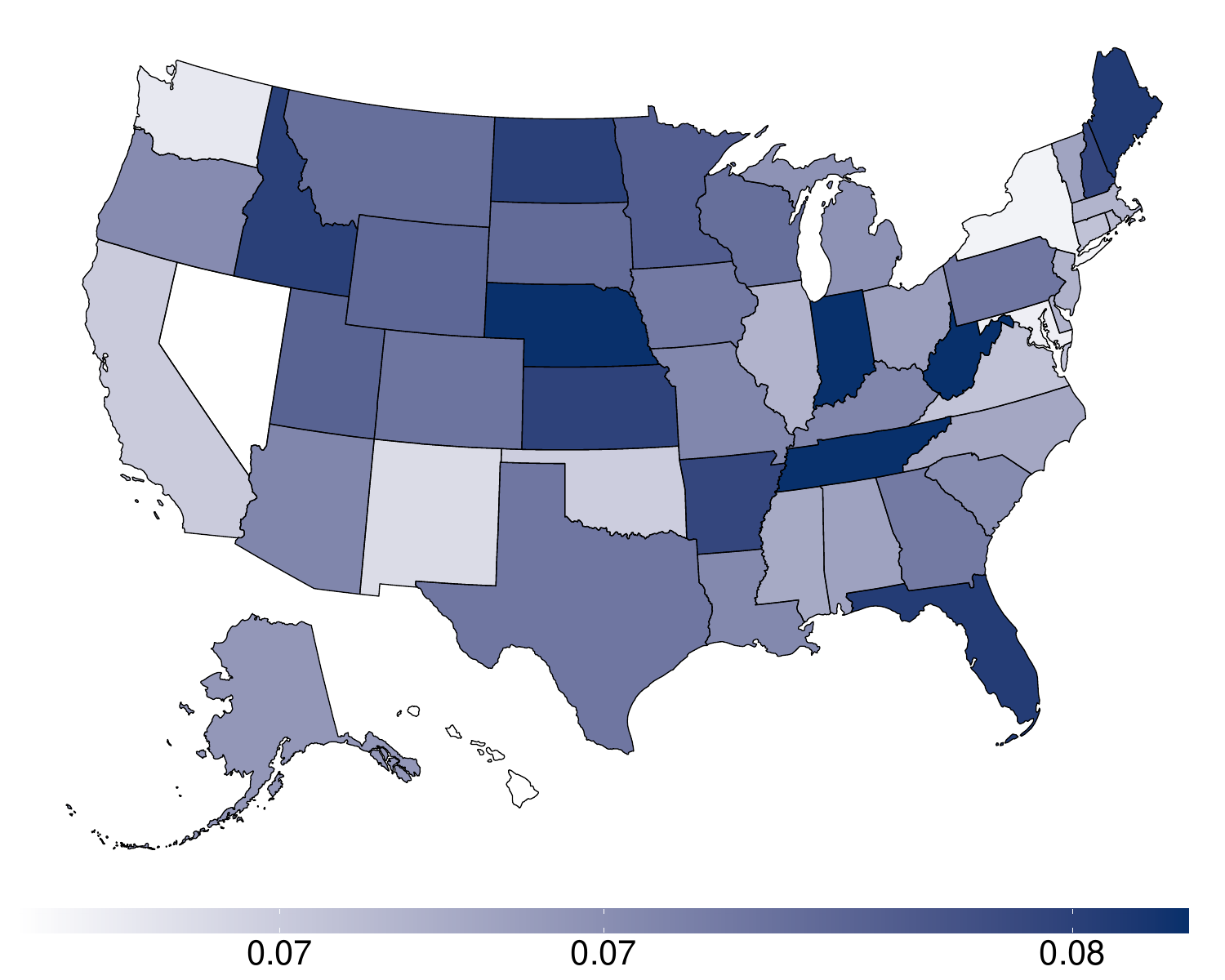}
        \caption{Model Population}
        \label{fig:MpopUSA}
    \end{subfigure}
         \hfill
    \begin{subfigure}{0.45\textwidth}
        \centering
        \includegraphics[width=\linewidth]{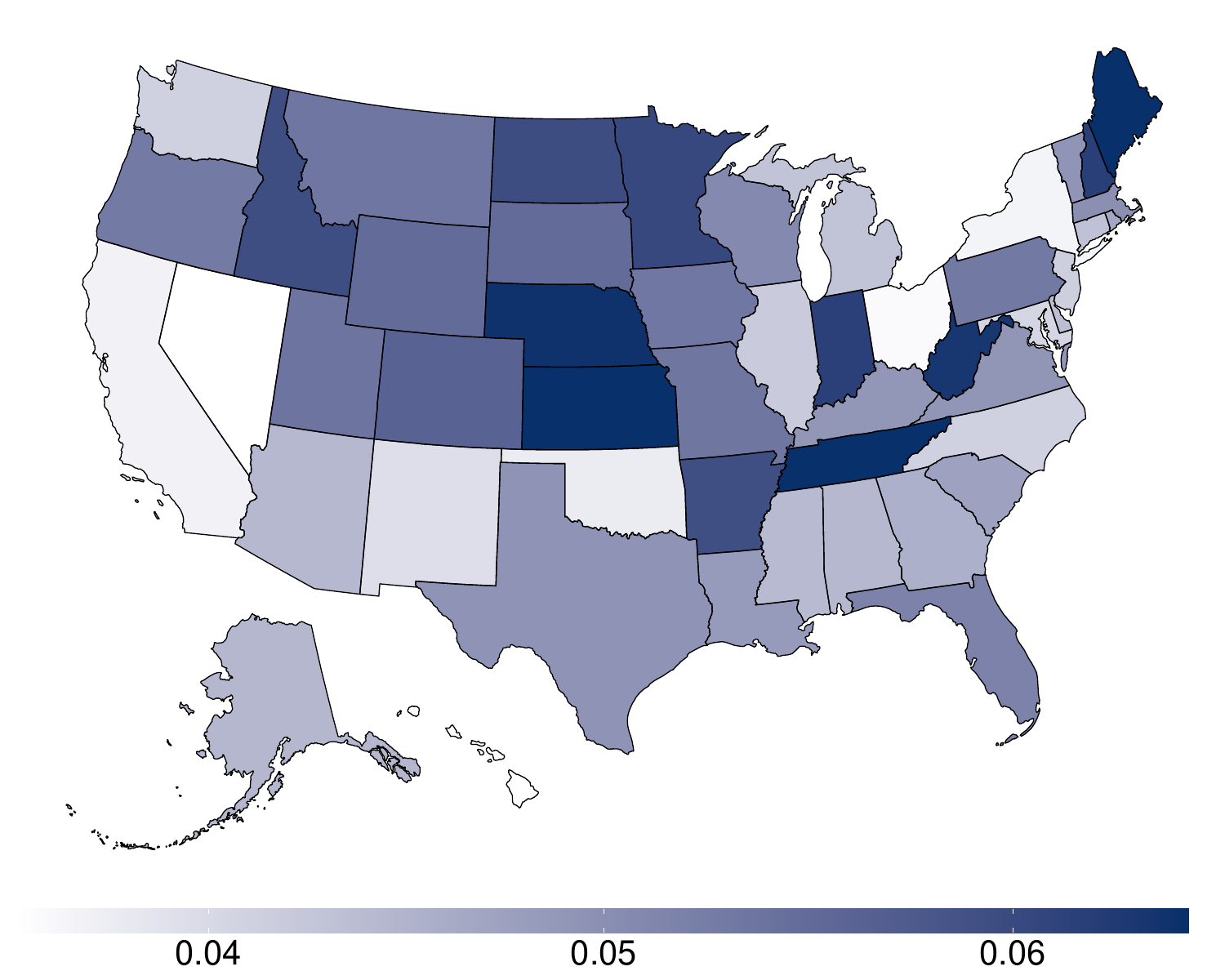}
        \caption{Model Balanced}
        \label{fig:MbalancedUSA}
    \end{subfigure}
    
    \caption{State-wide indices of fake news sharing. A darker colouring indicates that a state is likelier to share fake news. 
    The colouring is truncated $5^{th}$ to the $95^{th}$ percentile to limit the importance of outliers and enhance readability. The complete posterior distributions are available in the supplementary materials.}
    \label{fig:mainUSA}
\end{figure}

It is generally difficult to distinguish statistically between states in terms of the proportion of their population which is fake-news sharing. For inference based on Republican veracity assessments, Florida pops-out as a clear high-risk state. This is in line with findings from Figure \ref{fig:conint}, showing Republicans were more likely to label tweets related to Ron DeSantis as ``fake''. Figure \ref{fig:mainUSA} helps visualise the geographic anti-correlation between state-level estimates between based on partisan metrics. Using balanced and population metrics, we see a substantial degree of agreement in the ranking of the states -- it is for instance possible to say that the group of states characterised by the District of Columbia, Hawaii, Nevada, New York, and Washington are consistently assigned lower percentages of at-risk populations than Tennessee, Indiana, West Virginia and Nebraska. This tracks the partisan random effects and the `white\%' fixed effect, reinforcing the idea that red-states are -- though marginally given the overall small area-level heterogeneity -- more at-risk than blue-states. 

As a final point on the state-level estimates, we have generated tables with exact percentages and the expected number of people in each state of the US who shared fake news related to the pandemic during the period under study. This allows us to quantify, for the first time, the size of the fake news problem in terms of the exact number of at-risk people. Given the relatively low state-level heterogeneity, inference using absolute numbers suggests the more populous states have a greater absolute number of residents at risk of fake news sharing, ranging from $32,700$ individuals in California and $26,473$ in Texas, to a mere $422$ in the District of Columbia, and $521$ in Alaska. It is somewhat paradoxical that larger, more populated states would share the biggest burden of the fake news problem in absolute terms, when they tend to have populations which are on average more resistant to fake news sharing than those living in smaller, more sparsely populated states. 

\pagebreak

\section{Discussion} \label{discussion}
There is a gap in the democratic legitimacy of social-media content moderation, as a result of a loss of trust in expert-driven fact-checking. This paper seeks to address this gap by proposing an alternative detection approach which keeps humans in the loop and enhances the democratic value of fake news detection.

First, we have presented a methodology to aggregate crowd-sourced opinions on the veracity of digital content -- in our case, tweets related to the COVID-19 pandemic in the US. Our proposal here is that we can leverage the ``wisdom of the crowd'' to improve the democratic legitimacy of fake news detection.  We show how to compute a variety of comparable veracity metrics. These metrics highlight different attributes of the crowd's understanding of fake news; for example  $\nu^{\{...,\scriptsize \mbox{sample}\} }$ are natural metrics related to the assessments of the original sample of crowd-workers; $\nu^{\{...,\scriptsize \mbox{balanced}\} }$ are a set of metrics which privilege political balance over representativeness in deciding which pieces of digital content are to be considered fake; $\nu^{ \{\scriptsize \mbox{model}, \scriptsize \mbox{ pop.}\} }$ takes a ``one man, one vote'' approach, whereby the veracity scores of an exhaustive set of personae in the United States, for a given tweet, are predicted and accounted for, amounting to a ``wisdom of the representative crowd''; $\nu^{ \{...,\scriptsize \mbox{ party}\} }$ are partisan metrics which highlight heterogeneity in the definition of ``fake news'' within specific partisan strata of the population.

Two of our proposed metrics stand out as attractive for maximising democratic legitimacy, converge in terms of their inference about fake news sharing behaviour, and are likely to maximise the probability of correctly identifying fake news in absolute terms via the properties of the wisdom of the crowd \citep{murr2011wisdom}, and more generally the benefits of averaging heterogeneous predictions \citep{graefe2014combining,graefe2015improving}: $\nu^{\{\scriptsize \mbox{model}, \scriptsize \mbox{ pop.}\} }$ and $\nu^{\{\scriptsize \mbox{model}, \scriptsize \mbox{ balance}\} }$. These model-based metrics share some useful properties -- they both rely on an aggregation of the predicted veracity score for an exhaustive set of population personae. This is preferable to the sample-based naive aggregation approach, as there is an effort -- albeit likely biased to a degree that should be estimated in further work -- to generalise the learnings from the sample of fake news assessors to the population as a whole, including offline members of the polity. This is liable to increase both the accuracy of the metric via ensemble learning, as well as to improve democratic legitimacy by reducing reliance on silos or echo-chambers when making content moderation decisions. There is no obvious preference, to our mind, between the two metrics --  $\nu^{\{\scriptsize \mbox{model}, \scriptsize \mbox{ pop.}\} }$ is akin to a `popular vote' metric, whilst  $\nu^{\{\scriptsize \mbox{model}, \scriptsize \mbox{ balance}\} }$ is a stratified measure which distorts the popular vote in favour of equal representation. These correspond respectively to the \emph{Narrow} and \emph{Broad} conception of fairness in algorithmic decision making \citep{barocas2023fairness}: the `popular vote' approach focuses on individual fairness, counting every member of the polity as having the same amount of `say' in the content moderation decision; the balanced approach instead privileges a distortion of the popular vote in favour of increasing the weight of under-represented groups. In the specific case at hand, the balanced approach up-weights the views of Republicans relative to Democrats, but one could imagine veracity scores based on any kind of social cleavage beyond partisanship -- age, gender, ethnicity, rurality, religiosity, etc. -- which would be balanced using the same computations presented in this paper. 

\noindent The second objective of this paper was to understand the characteristics on individuals who shared fake news related to the pandemic on $\mathbb{X}$. Our findings can be summarised as follows: i. fake news sharing on pandemic-related topics between May and December $2022$ was a relatively rare phenomenon according to the wisdom of the crowd estimates (both population-based and balanced partisan metrics), with a baseline probability for the average individual consistently below $0.1$ (point estimate); ii. fake news sharing looks like a more serious problem from the lens of Democrats, with an average baseline probability of sharing fake news estimated at around $0.18$ when fitting a model to Democratic partisans veracity scores; iii. partisanship of the $\mathbb{X}$ user is the driving force in fake news sharing, with Democrats looking less-likely to share fake news relative to the average  $\mathbb{X}$; iv. exclusively when Republican partisans' definitions of fake news are used, it is Republicans who look less likely to share fake news; v. we find some evidence that women appear substantially less at-risk to share fake news than men; vi. we find some very weak evidence that $\mathbb{X}$ users over $40$ are at higher risk of sharing fake news; vii. state-level heterogeneity is contained but statistically meaningful. 





Our final effort was to generate a cross-states fake news sharing estimate, to provide a granular and comprehensive understanding of fake news dissemination across the United States. Despite the relatively contained state-level heterogeneity, we find a surprising degree of agreement and consistency across non-partisan veracity metrics in the ranking of states based on the \% and absolute number of people sharing fake news in each state. Driven by the partisanship effects, rural, predominantly white, ideologically conservative states are consistently ranked amongst the more at-risk states. Because of the limited effect that area-level contextual variables have on the estimates, the overall burden of the fake news sharing problem in terms of the absolute numbers of people at risk, falls broadly on the most populous states. 

\subsection{Limitations \& Further Research}

The conclusions of this paper need to be appropriately caveated. Firstly, with regards to inference about fake news sharing on $\mathbb{X}$, we note these findings are conditional on the specific set of tweets we have studied, and the context under which they were written and shared. These were tweets related to the pandemic, generated between May and December $2022$. The pandemic was a polarising event -- the strength of partisanship in this paper, both in terms of its ability to warp definitions of truth and affect spreading behaviour, is at least in part attributable to this. Moreover, the $2022$ congressional elections in the US happened during the period of study, further raising polarisation and the salience of partisanship in fake news sharing. The robustness of our findings should be further studied in different contexts. $\mathbb{X}$ as a platform has also changed dramatically since we began working on this paper, and it is possible the dynamics here presented no longer reflect the behaviour of the platform users due to unobservable selection effects. 

\noindent An important limitation of our methodology pertains to the Ordinal logistic regression model fit to the crowd-sourced ratings. In this model, we leveraged linear main effects and simple two-way interactions to capture systematic differences in veracity assessments due to worker characteristics and tweet topics. There is reason to believe the model we deployed is too simple to fully capture the complex character of these regularities. For starters, the ``topic'' of the tweet is a superficial measure of the characteristics of the tweet which interact with partisanship to distort veracity ratings. The Ron DeSantis example described in the paper illustrates that at the very least some form of sentiment by topic interaction is at play. A second related limitation of the ordinal logit model is that we only include interactions of tweet characteristics with partisanship, as opposed to other individual-level characteristics such as age or gender. Our model is parsimonious, focused primarily on our hypotheses around political polarisation. But allowing partisanship to affect the veracity ratings of topics differentially, and prohibiting age or gender from doing so, could be an important reason why we detect statistically significant anti-correlation amongst partisans, and not across other social cleavages. Future work could extend our step-1 model to include deep interaction and better account for complexity. 

\noindent Though this study uses Bayesian Hierarchical modelling for statistical inference, we do not propagate uncertainty at every stage, largely for computational reasons. Uncertainty about the veracity scores should be passed on to the second set of multilevel models, and be properly accounted for when estimating the effect of individual and state-level characteristics on fake news sharing. Uncertainty is further collapsed when we infer the demographic characteristics of online users. We have extracted these characteristics algorithmically, in non-generative fashion. The literature does however offer insights into estimating characteristics of social media users via Hierarchical Bayes \citep{barbera2015birds}. Future work should seek to develop a fully unified Bayesian model which incorporates uncertainty at three levels of analysis: i. the veracity assessments; ii. the demographic characteristics of online users; iii. the propensity to share fake news. The incorporation of this uncertainty can further contribute to improving the democratic legitimacy of fake news detection.

\pagebreak

\section*{Acknowledgements}
This paper was supported by a grant provided by the University Maastricht Behavioral Insights Center seeding grant provided by the School Of Business and Economics, Tongersestraat 53, 6211LM Maastricht, Netherlands.

\newpage

\bibliography{references.bib}

\end{document}


\doublespacing



\section{Veracity Estimation}

\scriptsize

\scriptsize
\begin{longtable}{p{4cm}rrrrrrrrr}
\caption{Summary table for the hierarchical ordinal logistic regression fit to the crowd-workers' veracity assessments.} \\
\hline

coefficient & mean & se\_mean & sd & 10\% & 50\% & 90\% & n\_eff & Rhat & $P(\beta > 0)$ \\ 
 \hline
\endfirsthead

\hline

coefficient & mean & se\_mean & sd & 10\% & 50\% & 90\% & n\_eff & Rhat & $P(\beta > 0)$ \\ 
 \hline
\endhead

  \hline
 Theta[1] & -1.38 & 0.07 & 1.24 & -2.95 & -1.36 & 0.07 & 352.14 & 0.99 & 0.11\\ 
  Theta[2] & -0.09 & 0.07 & 1.24 & -1.64 & -0.06 & 1.37 & 352.42 & 0.99 & 0.49\\ 
  Theta[3] & 1.62 & 0.07 & 1.24 & 0.06 & 1.66 & 3.08 & 352.81 & 0.99 & 0.90\\ 
  Intercept & -0.03 & 0.05 & 1.01 & -1.37 & -0.09 & 1.24 & 347.10 & 1.00 & 0.47\\ 
  Party: Democrat & 0.60 & 0.03 & 0.67 & -0.24 & 0.58 & 1.45 & 409.76 & 1.00& 0.81 \\ 
  Party: Neutral & -0.70 & 0.04 & 0.72 & -1.67 & -0.65 & 0.14 & 329.89 & 1.00 & 0.15 \\ 
  Party: Republican & 0.07 & 0.03 & 0.57 & -0.64 & 0.06 & 0.83 & 398.27 & 0.99 & 0.53 \\ 
  $\sigma_{Party}$ & 0.33 & 0.01 & 0.27 & 0.11 & 0.25 & 0.64 & 363.53 & 1.00 & 1.00\\
\end{longtable}

\newpage
\section{Fake News Sharing Inference\\Using the 10\% Threshold}

\scriptsize
\centering
\begin{longtable}{rrrrrrrrrrr}
\caption{Summary of the posterior distribution for the parameters of the model predicting ``likelihood to spread fake news'' using the aggregated ratings of Democratic survey participants - $\nu^{ \{ \mbox{\scriptsize naive}, \mbox{ \scriptsize party} \} }_{j=D}$.} \\

\hline
coefficient & mean & se\_mean & sd & 10\% & 50\% & 90\% & n\_eff & Rhat & $P(\beta >0)$\\ 
  \hline
\endfirsthead
\hline
coefficient & mean & se\_mean & sd & 10\% & 50\% & 90\% & n\_eff & Rhat & $P(\beta >0)$\\ 
\hline
\endhead
\hline
\endfoot
\hline
\endlastfoot
intercept & -1.63 & 0.01 & 0.59 & -2.29 & -1.73 & -0.83 & 2310.03 & 1.00 & 0.02 \\ 
  Sex:Female & -0.11 & 0.00 & 0.11 & -0.25 & -0.10 & 0.04 & 2869.69 & 1.00 & 0.17 \\ 
  sigma Age & 0.12 & 0.00 & 0.15 & 0.01 & 0.08 & 0.28 & 2348.63 & 1.00 & 1.00 \\ 
  Age:-18 & 0.03 & 0.00 & 0.12 & -0.07 & 0.01 & 0.16 & 2429.90 & 1.00 & 0.63 \\ 
  Age:19-29 & -0.01 & 0.00 & 0.11 & -0.12 & -0.00 & 0.09 & 2432.12 & 1.00 & 0.48 \\ 
  Age:30-39 & -0.04 & 0.00 & 0.12 & -0.17 & -0.01 & 0.06 & 2328.11 & 1.00 & 0.40 \\ 
  Age:40+ & -0.03 & 0.00 & 0.11 & -0.14 & -0.01 & 0.06 & 2538.43 & 1.00 & 0.39 \\ 
  sigma State & 0.11 & 0.00 & 0.08 & 0.02 & 0.10 & 0.21 & 2111.61 & 1.00 & 1.00 \\
\end{longtable}

\newpage
\scriptsize
\centering
\begin{longtable}{rrrrrrrrrr}
\caption{Summary of the posterior distribution for the parameters of the model predicting ``likelihood to spread fake news'' using the aggregated ratings of Republican survey participants - $\nu^{ \{ \mbox{\scriptsize naive}, \mbox{ \scriptsize party} \} }_{j=R}$.}  \\

\hline
 coefficient& mean & se\_mean & sd & 10\% & 50\% & 90\% & n\_eff & Rhat & $P(\beta >0)$ \\ 
  \hline
\endfirsthead
\hline
coefficient & mean & se\_mean & sd & 10\% & 50\% & 90\% & n\_eff & Rhat & $P(\beta >0)$ \\ 
\hline
\endhead
\hline
\endfoot
\hline
\endlastfoot
intercept & -1.88 & 0.01 & 0.45 & -2.33 & -1.98 & -1.28 & 967.78 & 1.01 & 0.00 \\ 
  Sex:Female & -0.02 & 0.00 & 0.10 & -0.14 & -0.02 & 0.11 & 5261.16 & 1.00 & 0.44 \\ 
  sigma Age & 0.18 & 0.01 & 0.21 & 0.02 & 0.12 & 0.38 & 622.49 & 1.02 & 1.00 \\ 
  Age:-18 & 0.05 & 0.01 & 0.17 & -0.07 & 0.03 & 0.24 & 833.52 & 1.01 & 0.67 \\ 
  Age:19-29 & -0.06 & 0.01 & 0.17 & -0.23 & -0.02 & 0.06 & 625.62 & 1.01 & 0.35 \\ 
  Age:30-39 & -0.10 & 0.01 & 0.18 & -0.29 & -0.05 & 0.04 & 686.23 & 1.01 & 0.25 \\ 
  Age:40+ & -0.02 & 0.01 & 0.16 & -0.15 & 0.00 & 0.11 & 588.72 & 1.01 & 0.51 \\ 
  sigma State & 0.13 & 0.00 & 0.09 & 0.02 & 0.12 & 0.25 & 4526.55 & 1.00 & 1.00 \\
\end{longtable}

\newpage
\scriptsize
\centering
\begin{longtable}{rrrrrrrrrr}
\caption{Summary of the posterior distribution for the parameters of the model predicting ``likelihood to spread fake news'' using the average survey rating - $\nu^{ \{ \mbox{\scriptsize naive}, \mbox{ \scriptsize sample} \} }$.}  \\

\hline
coefficient & mean & se\_mean & sd & 10\% & 50\% & 90\% & n\_eff & Rhat &\\ 
  \hline
\endfirsthead
\hline
coefficient & mean & se\_mean & sd & 10\% & 50\% & 90\% & n\_eff & Rhat &\\ 
\hline
\endhead
\hline
\endfoot
\hline
\endlastfoot
intercept & -1.78 & 0.01 & 0.21 & -1.98 & -1.80 & -1.53 & 1337.73 & 1.01 & 0.00 \\ 
  Sex:Female & -0.12 & 0.00 & 0.09 & -0.24 & -0.12 & -0.01 & 2279.21 & 1.00 & 0.08 \\ 
  sigma Age & 0.12 & 0.00 & 0.14 & 0.01 & 0.08 & 0.27 & 1098.55 & 1.00 & 1.00 \\ 
  Age:-18 & 0.01 & 0.00 & 0.12 & -0.10 & 0.00 & 0.12 & 1335.57 & 1.00 & 0.55 \\ 
  Age:19-29 & 0.02 & 0.00 & 0.10 & -0.06 & 0.01 & 0.13 & 1115.84 & 1.00 & 0.63 \\ 
  Age:30-39 & -0.03 & 0.00 & 0.11 & -0.15 & -0.01 & 0.06 & 1231.90 & 1.00 & 0.40 \\ 
  Age:40+ & -0.05 & 0.00 & 0.11 & -0.16 & -0.02 & 0.04 & 1053.89 & 1.00 & 0.29 \\ 
  sigma State & 0.14 & 0.00 & 0.08 & 0.04 & 0.13 & 0.24 & 2154.07 & 1.00 & 1.00 \\
\end{longtable}

\newpage
\scriptsize
\centering
\begin{longtable}{rrrrrrrrrr}
\caption{Summary of the posterior distribution for the parameters of the model predicting ``likelihood to spread fake news'' using as many Republicans as Democrats per tweets - $\nu^{ \{ \mbox{\scriptsize naive}, \mbox{ \scriptsize balanced} \} }$.}  \\

\hline
coefficient & mean & se\_mean & sd & 10\% & 50\% & 90\% & n\_eff & Rhat & $P(\beta >0)$ \\ 
  \hline
\endfirsthead
\hline
coefficient & mean & se\_mean & sd & 10\% & 50\% & 90\% & n\_eff & Rhat & $P(\beta >0)$ \\ 
\hline
\endhead
\hline
\endfoot
\hline
\endlastfoot
intercept & -2.00 & 0.01 & 0.28 & -2.28 & -2.05 & -1.63 & 1752.86 & 1.01 & 0.00 \\ 
  Sex:Female & -0.08 & 0.00 & 0.11 & -0.22 & -0.08 & 0.06 & 6700.61 & 1.00 & 0.23 \\ 
  sigma Age & 0.21 & 0.00 & 0.21 & 0.03 & 0.15 & 0.48 & 2527.07 & 1.00 & 1.00 \\ 
  Age:-18 & 0.08 & 0.00 & 0.18 & -0.08 & 0.05 & 0.30 & 2249.97 & 1.00 & 0.71 \\ 
  Age:19-29 & -0.13 & 0.01 & 0.20 & -0.37 & -0.08 & 0.03 & 1312.19 & 1.00 & 0.22 \\ 
  Age:30-39 & -0.03 & 0.00 & 0.18 & -0.22 & -0.00 & 0.13 & 1284.55 & 1.00 & 0.49 \\ 
  Age:40+ & -0.09 & 0.01 & 0.18 & -0.29 & -0.04 & 0.06 & 1097.17 & 1.00 & 0.30 \\ 
  sigma State & 0.07 & 0.00 & 0.06 & 0.01 & 0.06 & 0.15 & 6813.15 & 1.00 & 1.00 \\
\end{longtable}

\newpage
\scriptsize
\centering
\begin{longtable}{rrrrrrrrrr}
\caption{Summary of the posterior distribution for the parameters of the model predicting ``likelihood to spread fake news'' using the weighted assessment of the survey participants - $\nu^{ \{ \mbox{\scriptsize naive}, \mbox{ \scriptsize pop.} \} }.$}  \\

\hline
coefficient & mean & se\_mean & sd & 10\% & 50\% & 90\% & n\_eff & Rhat &\\ 
  \hline
\endfirsthead
\hline
coefficient & mean & se\_mean & sd & 10\% & 50\% & 90\% & n\_eff & Rhat &\\ 
\hline
\endhead
\hline
\endfoot
\hline
\endlastfoot
intercept & -2.08 & 0.02 & 0.27 & -2.32 & -2.13 & -1.76 & 202.81 & 1.02 & 0.00 \\ 
  Sex:Female & -0.07 & 0.00 & 0.10 & -0.20 & -0.07 & 0.06 & 1960.27 & 1.00 & 0.25 \\ 
  sigma Age & 0.13 & 0.00 & 0.15 & 0.02 & 0.08 & 0.28 & 2202.80 & 1.00 & 1.00 \\ 
  Age:-18 & 0.02 & 0.00 & 0.12 & -0.08 & 0.01 & 0.15 & 2400.85 & 1.00 & 0.60 \\ 
  Age:19-29 & -0.00 & 0.00 & 0.10 & -0.10 & 0.00 & 0.10 & 2557.10 & 1.00 & 0.54 \\ 
  Age:30-39 & -0.05 & 0.00 & 0.12 & -0.18 & -0.02 & 0.05 & 1935.03 & 1.00 & 0.34 \\ 
  Age:40+ & -0.03 & 0.00 & 0.10 & -0.15 & -0.01 & 0.06 & 2358.32 & 1.00 & 0.37 \\ 
  sigma State & 0.17 & 0.00 & 0.09 & 0.05 & 0.17 & 0.29 & 1294.72 & 1.00 & 1.00 \\
\end{longtable}

\newpage
\scriptsize
\centering
\begin{longtable}{rrrrrrrrrr}
\caption{Summary of the posterior distribution for the parameters of the model predicting ``likelihood to spread fake news''  using the model-based ratings of Democrat personas - $\nu^{ \{ \mbox{\scriptsize model}, \mbox{ \scriptsize party} \} }_{j=D}$.}  \\

\hline
coefficient & mean & se\_mean & sd & 10\% & 50\% & 90\% & n\_eff & Rhat & $P(\beta >0)$ \\ 
  \hline
\endfirsthead
\hline
coefficient & mean & se\_mean & sd & 10\% & 50\% & 90\% & n\_eff & Rhat & $P(\beta >0)$ \\ 
\hline
\endhead
\hline
\endfoot
\hline
\endlastfoot
intercept & -1.83 & 0.02 & 0.49 & -2.33 & -1.93 & -1.17 & 775.00 & 1.00 & 0.00 \\ 
  Sex:Female & -0.05 & 0.00 & 0.10 & -0.18 & -0.05 & 0.08 & 2426.08 & 1.00 & 0.29 \\ 
  sigma Age & 0.10 & 0.00 & 0.13 & 0.01 & 0.06 & 0.23 & 2330.63 & 1.00 & 1.00 \\ 
  Age:-18 & -0.02 & 0.00 & 0.11 & -0.12 & -0.00 & 0.08 & 2443.96 & 1.00 & 0.46 \\ 
  Age:19-29 & -0.01 & 0.00 & 0.09 & -0.10 & 0.00 & 0.08 & 2320.61 & 1.00 & 0.50 \\ 
  Age:30-39 & -0.01 & 0.00 & 0.10 & -0.11 & -0.00 & 0.07 & 2487.28 & 1.00 & 0.45 \\ 
  Age:40+ & -0.02 & 0.00 & 0.09 & -0.11 & -0.01 & 0.05 & 2390.02 & 1.00 & 0.40 \\ 
  sigma State & 0.07 & 0.00 & 0.05 & 0.01 & 0.06 & 0.15 & 2550.13 & 1.00 & 1.00 \\
\end{longtable}

\newpage
\scriptsize
\centering
\begin{longtable}{rrrrrrrrrr}
\caption{Summary of the posterior distribution for the parameters of the model predicting ``likelihood to spread fake news'' using the model-based ratings of Republican personas - $\nu^{ \{ \mbox{\scriptsize model}, \mbox{ \scriptsize party} \} }_{j=R}$.} \\

\hline
coefficient & mean & se\_mean & sd & 10\% & 50\% & 90\% & n\_eff & Rhat &\\ 
  \hline
\endfirsthead
\hline
 coefficient& mean & se\_mean & sd & 10\% & 50\% & 90\% & n\_eff & Rhat &\\ 
\hline
\endhead
\hline
\endfoot
\hline
\endlastfoot
intercept & -2.16 & 0.01 & 0.36 & -2.54 & -2.23 & -1.66 & 1166.12 & 1.01 & 0.00 \\ 
  Sex:Female & -0.16 & 0.00 & 0.10 & -0.29 & -0.16 & -0.04 & 2112.84 & 1.00 & 0.05 \\ 
  sigma Age & 0.29 & 0.01 & 0.25 & 0.10 & 0.22 & 0.57 & 1124.15 & 1.00 & 1.00 \\ 
  Age:-18 & -0.08 & 0.01 & 0.24 & -0.35 & -0.04 & 0.16 & 1155.44 & 1.01 & 0.38 \\ 
  Age:19-29 & -0.12 & 0.01 & 0.22 & -0.35 & -0.08 & 0.08 & 893.07 & 1.01 & 0.26 \\ 
  Age:30-39 & -0.18 & 0.01 & 0.23 & -0.44 & -0.15 & 0.03 & 954.39 & 1.01 & 0.15 \\ 
  Age:40+ & 0.11 & 0.01 & 0.21 & -0.08 & 0.13 & 0.31 & 926.86 & 1.01 & 0.82 \\ 
  sigma State & 0.32 & 0.00 & 0.06 & 0.24 & 0.31 & 0.40 & 2229.80 & 1.00 & 1.00 \\
\end{longtable}

\newpage
\scriptsize
\centering
\begin{longtable}{rrrrrrrrrr}
\caption{Summary of the posterior distribution for the parameters of the model predicting ``likelihood to spread fake news'' using the model-based ratings of the survey participants -  $\nu^{ \{ \mbox{\scriptsize model}, \mbox{ \scriptsize sample} \} }$.}  \\

\hline
 coefficient& mean & se\_mean & sd & 10\% & 50\% & 90\% & n\_eff & Rhat &\\ 
  \hline
\endfirsthead
\hline
coefficient & mean & se\_mean & sd & 10\% & 50\% & 90\% & n\_eff & Rhat &\\ 
\hline
\endhead
\hline
\endfoot
\hline
\endlastfoot
intercept & -2.04 & 0.02 & 0.36 & -2.40 & -2.10 & -1.56 & 508.00 & 1.02 & 0.00 \\ 
  Sex:Female & 0.06 & 0.00 & 0.10 & -0.06 & 0.07 & 0.19 & 2434.47 & 1.00 & 0.74 \\ 
  sigma Age & 0.23 & 0.00 & 0.20 & 0.04 & 0.18 & 0.49 & 1808.94 & 1.00 & 1.00 \\ 
  Age:-18 & 0.08 & 0.00 & 0.19 & -0.10 & 0.06 & 0.30 & 2216.27 & 1.00 & 0.70 \\ 
  Age:19-29 & -0.10 & 0.00 & 0.18 & -0.32 & -0.06 & 0.07 & 2245.35 & 1.00 & 0.27 \\ 
  Age:30-39 & -0.16 & 0.00 & 0.20 & -0.42 & -0.12 & 0.02 & 2137.78 & 1.00 & 0.17 \\ 
  Age:40+ & 0.02 & 0.00 & 0.17 & -0.16 & 0.03 & 0.19 & 2204.13 & 1.00 & 0.61 \\ 
  sigma State & 0.16 & 0.00 & 0.08 & 0.06 & 0.16 & 0.26 & 1824.83 & 1.00 & 1.00 \\
\end{longtable}

\newpage
\scriptsize
\centering
\begin{longtable}{rrrrrrrrrr}
\caption{Summary of the posterior distribution for the parameters of the model predicting ``likelihood to spread fake news'' using the model-based ratings of Democrat and Republican personas - $\nu^{ \{ \mbox{\scriptsize model}, \mbox{ \scriptsize balanced} \} }$.}  \\

\hline
coefficient & mean & se\_mean & sd & 10\% & 50\% & 90\% & n\_eff & Rhat &\\ 
  \hline
\endfirsthead
\hline
coefficient & mean & se\_mean & sd & 10\% & 50\% & 90\% & n\_eff & Rhat &\\ 
\hline
\endhead
\hline
\endfoot
\hline
\endlastfoot
intercept & -2.10 & 0.01 & 0.27 & -2.39 & -2.15 & -1.73 & 977.15 & 1.01 & 0.00 \\ 
  Sex:Female & -0.17 & 0.00 & 0.10 & -0.30 & -0.17 & -0.04 & 2100.12 & 1.00 & 0.05 \\ 
  sigma Age & 0.20 & 0.00 & 0.17 & 0.04 & 0.15 & 0.41 & 2177.37 & 1.00 & 1.00 \\ 
  Age:-18 & -0.06 & 0.00 & 0.17 & -0.26 & -0.03 & 0.10 & 2428.57 & 1.00 & 0.38 \\ 
  Age:19-29 & -0.08 & 0.00 & 0.15 & -0.26 & -0.05 & 0.06 & 2399.03 & 1.00 & 0.28 \\ 
  Age:30-39 & -0.08 & 0.00 & 0.16 & -0.26 & -0.05 & 0.07 & 2213.66 & 1.00 & 0.29 \\ 
  Age:40+ & 0.08 & 0.00 & 0.14 & -0.05 & 0.08 & 0.24 & 2434.36 & 1.00 & 0.79 \\ 
  sigma State & 0.23 & 0.00 & 0.08 & 0.13 & 0.22 & 0.33 & 1938.65 & 1.01 & 1.00 \\
\end{longtable}

\newpage
\scriptsize
\centering
\begin{longtable}{rrrrrrrrrr}
\caption{Summary of the posterior distribution for the parameters of the model predicting ``likelihood to spread fake news'' using the wisdom of the synthetic crowds - $\nu^{ \{ \mbox{\scriptsize model}, \mbox{ \scriptsize pop.} \} }.$}  \\

\hline
coefficient & mean & se\_mean & sd & 10\% & 50\% & 90\% & n\_eff & Rhat & $P(\beta >0)$ \\ 
  \hline
\endfirsthead
\hline
 coefficient& mean & se\_mean & sd & 10\% & 50\% & 90\% & n\_eff & Rhat &\\ 
\hline
\endhead
\hline
\endfoot
\hline
\endlastfoot
intercept & -2.10 & 0.01 & 0.29 & -2.41 & -2.14 & -1.71 & 1812.04 & 1.00 & 0.00 \\ 
  Sex:Female & -0.15 & 0.00 & 0.10 & -0.28 & -0.15 & -0.02 & 2403.18 & 1.00 & 0.06 \\ 
  sigma Age & 0.23 & 0.00 & 0.20 & 0.05 & 0.17 & 0.48 & 1883.24 & 1.00 & 1.00 \\ 
  Age:-18 & -0.08 & 0.00 & 0.20 & -0.33 & -0.04 & 0.10 & 1964.47 & 1.00 & 0.33 \\ 
  Age:19-29 & -0.09 & 0.00 & 0.18 & -0.29 & -0.06 & 0.08 & 1653.03 & 1.00 & 0.28 \\ 
  Age:30-39 & -0.09 & 0.00 & 0.18 & -0.29 & -0.06 & 0.08 & 1685.60 & 1.00 & 0.28 \\ 
  Age:40+ & 0.09 & 0.00 & 0.17 & -0.05 & 0.09 & 0.27 & 1724.51 & 1.00 & 0.80 \\ 
  sigma State & 0.22 & 0.00 & 0.09 & 0.12 & 0.22 & 0.33 & 1930.10 & 1.00 & 1.00 \\
\end{longtable}

\newpage

\section{Fake News Sharing Inference\\Using the \textit{Not Very Accurate} Threshold}

\scriptsize
\centering
\begin{longtable}{rrrrrrrrrr}
\caption{Summary of the posterior distribution for the parameters of the model predicting ``likelihood to spread fake news'' using the aggregated ratings of Democratic survey participants - $\nu^{ \{ \mbox{\scriptsize naive}, \mbox{ \scriptsize party} \} }_{j=D}$. Fake news is defined as news that is rated at or below \it{Not Very Accurate}.}  \\

\hline
coefficient & mean & se\_mean & sd & 10\% & 50\% & 90\% & n\_eff & Rhat &$P(\beta >0)$\\ 
  \hline
\endfirsthead
\hline
coefficient & mean & se\_mean & sd & 10\% & 50\% & 90\% & n\_eff & Rhat &$P(\beta >0)$\\ 
\hline
\endhead
\hline
\endfoot
\hline
\endlastfoot
intercept & -0.73 & 0.01 & 0.43 & -1.22 & -0.77 & -0.19 & 4482.91 & 1.00 & 0.05 \\ 
  Sex:Female & -0.06 & 0.00 & 0.07 & -0.16 & -0.07 & 0.03 & 8000.26 & 1.00 & 0.18 \\ 
  sigma Age & 0.09 & 0.00 & 0.11 & 0.01 & 0.06 & 0.21 & 7079.27 & 1.00 & 1.00 \\ 
  Age:-18 & 0.03 & 0.00 & 0.09 & -0.05 & 0.01 & 0.13 & 7305.41 & 1.00 & 0.63 \\ 
  Age:19-29 & -0.01 & 0.00 & 0.08 & -0.10 & -0.00 & 0.06 & 7471.04 & 1.00 & 0.45 \\ 
  Age:30-39 & 0.01 & 0.00 & 0.08 & -0.07 & 0.00 & 0.09 & 6583.03 & 1.00 & 0.57 \\ 
  Age:40+ & -0.03 & 0.00 & 0.08 & -0.12 & -0.02 & 0.03 & 6554.74 & 1.00 & 0.31 \\ 
  sigma State & 0.05 & 0.00 & 0.04 & 0.01 & 0.04 & 0.10 & 7464.21 & 1.00 & 1.00 \\
\end{longtable}

\newpage
\scriptsize
\centering
\begin{longtable}{rrrrrrrrrr}
\caption{Summary of the posterior distribution for the parameters of the model predicting ``likelihood to spread fake news'' using the aggregated ratings of Republican survey participants - $\nu^{ \{ \mbox{\scriptsize naive}, \mbox{ \scriptsize party} \} }_{j=R}$. Fake news is defined as news that is rated at or below \it{Not Very Accurate}.}  \\

\hline
 coefficient& mean & se\_mean & sd & 10\% & 50\% & 90\% & n\_eff & Rhat & $P(\beta >0)$ \\ 
  \hline
\endfirsthead
\hline
coefficient & mean & se\_mean & sd & 10\% & 50\% & 90\% & n\_eff & Rhat & $P(\beta >0)$ \\ 
\hline
\endhead
\hline
\endfoot
\hline
\endlastfoot
intercept & -0.78 & 0.01 & 0.34 & -1.16 & -0.80 & -0.37 & 3683.88 & 1.00 & 0.03 \\ 
  Sex:Female & 0.09 & 0.00 & 0.07 & 0.00 & 0.09 & 0.17 & 7252.99 & 1.00 & 0.90 \\ 
  sigma Age & 0.09 & 0.00 & 0.11 & 0.01 & 0.05 & 0.19 & 7114.66 & 1.00 & 1.00 \\ 
  Age:-18 & 0.02 & 0.00 & 0.08 & -0.05 & 0.01 & 0.12 & 7111.64 & 1.00 & 0.62 \\ 
  Age:19-29 & -0.03 & 0.00 & 0.08 & -0.11 & -0.01 & 0.04 & 7903.80 & 1.00 & 0.37 \\ 
  Age:30-39 & -0.02 & 0.00 & 0.08 & -0.10 & -0.01 & 0.05 & 7642.44 & 1.00 & 0.41 \\ 
  Age:40+ & 0.01 & 0.00 & 0.07 & -0.06 & 0.00 & 0.07 & 7391.30 & 1.00 & 0.56 \\ 
  sigma State & 0.07 & 0.00 & 0.05 & 0.01 & 0.06 & 0.14 & 6583.46 & 1.00 & 1.00 \\
\end{longtable}

\newpage
\scriptsize
\centering
\begin{longtable}{rrrrrrrrrr}
\caption{Summary of the posterior distribution for the parameters of the model predicting ``likelihood to spread fake news'' using the average survey rating - $\nu^{ \{ \mbox{\scriptsize naive}, \mbox{ \scriptsize sample} \} }$. Fake news is defined as news that is rated at or below \it{Not Very Accurate}.}  \\

\hline
coefficient & mean & se\_mean & sd & 10\% & 50\% & 90\% & n\_eff & Rhat &$P(\beta >0)$\\  
  \hline
\endfirsthead
\hline
coefficient & mean & se\_mean & sd & 10\% & 50\% & 90\% & n\_eff & Rhat & $P(\beta >0)$ \\ 
\hline
\endhead
\hline
\endfoot
\hline
\endlastfoot

intercept & -1.78 & 0.01 & 0.22 & -2.00 & -1.81 & -1.50 & 706.25 & 1.01 & 0.00 \\ 
  Sex:Female & -0.13 & 0.00 & 0.09 & -0.24 & -0.12 & -0.01 & 7815.71 & 1.00 & 0.08 \\ 
  sigma Age & 0.12 & 0.01 & 0.15 & 0.01 & 0.08 & 0.27 & 702.73 & 1.02 & 1.00 \\ 
  Age:-18 & 0.01 & 0.00 & 0.11 & -0.10 & 0.00 & 0.12 & 758.55 & 1.02 & 0.54 \\ 
  Age:19-29 & 0.02 & 0.00 & 0.11 & -0.07 & 0.01 & 0.13 & 664.50 & 1.02 & 0.63 \\ 
  Age:30-39 & -0.03 & 0.00 & 0.11 & -0.15 & -0.01 & 0.06 & 560.82 & 1.02 & 0.40 \\ 
  Age:40+ & -0.05 & 0.00 & 0.11 & -0.16 & -0.02 & 0.04 & 514.30 & 1.02 & 0.29 \\ 
  sigma State & 0.14 & 0.00 & 0.08 & 0.04 & 0.14 & 0.24 & 4965.78 & 1.00 & 1.00 \\
\end{longtable}

\newpage
\scriptsize
\centering
\begin{longtable}{rrrrrrrrrr}
\caption{Summary of the posterior distribution for the parameters of the model predicting ``likelihood to spread fake news'' using as many Republicans as Democrats per tweet - $\nu^{ \{ \mbox{\scriptsize naive}, \mbox{ \scriptsize balanced} \} }$. Fake news is defined as news that is rated at or below \it{Not Very Accurate}.}  \\

\hline
coefficient & mean & se\_mean & sd & 10\% & 50\% & 90\% & n\_eff & Rhat & $P(\beta >0)$ \\ 
  \hline
\endfirsthead
\hline
coefficient & mean & se\_mean & sd & 10\% & 50\% & 90\% & n\_eff & Rhat & $P(\beta >0)$ \\ 
\hline
\endhead
\hline
\endfoot
\hline
\endlastfoot
intercept & -1.55 & 0.02 & 0.24 & -1.79 & -1.59 & -1.25 & 199.36 & 1.04 & 0.00 \\ 
  Sex:Female & -0.13 & 0.00 & 0.09 & -0.25 & -0.13 & -0.01 & 3470.25 & 1.00 & 0.08 \\ 
  sigma Age & 0.20 & 0.01 & 0.21 & 0.03 & 0.14 & 0.45 & 496.96 & 1.02 & 1.00 \\ 
  Age:-18 & 0.10 & 0.01 & 0.16 & -0.04 & 0.07 & 0.31 & 1043.35 & 1.01 & 0.77 \\ 
  Age:19-29 & -0.10 & 0.01 & 0.17 & -0.31 & -0.06 & 0.04 & 296.52 & 1.03 & 0.25 \\ 
  Age:30-39 & -0.06 & 0.01 & 0.16 & -0.25 & -0.03 & 0.08 & 364.65 & 1.03 & 0.35 \\ 
  Age:40+ & -0.06 & 0.01 & 0.16 & -0.24 & -0.02 & 0.08 & 244.09 & 1.04 & 0.37 \\ 
  sigma State & 0.06 & 0.00 & 0.05 & 0.01 & 0.05 & 0.12 & 4991.42 & 1.00 & 1.00 \\
\end{longtable}

\newpage
\scriptsize
\centering
\begin{longtable}{rrrrrrrrrr}
\caption{Summary of the posterior distribution for the parameters of the model predicting ``likelihood to spread fake news'' using the weighted assessment of the survey participants - $\nu^{ \{ \mbox{\scriptsize naive}, \mbox{ \scriptsize pop.} \} }$. Fake news is defined as news that is rated at or below \it{Not Very Accurate}.} \\

\hline
coefficient & mean & se\_mean & sd & 10\% & 50\% & 90\% & n\_eff & Rhat & $P(\beta >0)$ \\ 
  \hline
\endfirsthead
\hline
coefficient & mean & se\_mean & sd & 10\% & 50\% & 90\% & n\_eff & Rhat & $P(\beta >0)$\\ 
\hline
\endhead
\hline
\endfoot
\hline
\endlastfoot
intercept & -2.00 & 0.01 & 0.26 & -2.27 & -2.05 & -1.65 & 527.63 & 1.02 & 0.00 \\ 
  Sex:Female & -0.04 & 0.00 & 0.09 & -0.16 & -0.04 & 0.08 & 6885.45 & 1.00 & 0.32 \\ 
  sigma Age & 0.15 & 0.01 & 0.17 & 0.02 & 0.09 & 0.35 & 334.15 & 1.03 & 1.00 \\ 
  Age:-18 & 0.03 & 0.01 & 0.16 & -0.08 & 0.02 & 0.19 & 174.79 & 1.06 & 0.63 \\ 
  Age:19-29 & -0.02 & 0.01 & 0.16 & -0.15 & -0.00 & 0.10 & 124.75 & 1.09 & 0.49 \\ 
  Age:30-39 & -0.08 & 0.01 & 0.17 & -0.26 & -0.03 & 0.03 & 139.92 & 1.08 & 0.28 \\ 
  Age:40+ & -0.04 & 0.01 & 0.16 & -0.17 & -0.01 & 0.07 & 123.31 & 1.09 & 0.41 \\ 
  sigma State & 0.20 & 0.00 & 0.09 & 0.08 & 0.20 & 0.32 & 2450.65 & 1.01 & 1.00 \\
\end{longtable}

\newpage
\scriptsize
\centering 
\begin{longtable}{rrrrrrrrrr}
\caption{Summary of the posterior distribution for the parameters of the model predicting ``likelihood to spread fake news'' using the model-based ratings of Democrat personas - $\nu^{ \{ \mbox{\scriptsize model}, \mbox{ \scriptsize party} \} }_{j=D}$. Fake news is defined as news that is rated at or below \it{Not Very Accurate}.} \\

\hline
coefficient & mean & se\_mean & sd & 10\% & 50\% & 90\% & n\_eff & Rhat &\\ 
  \hline
\endfirsthead
\hline
coefficient & mean & se\_mean & sd & 10\% & 50\% & 90\% & n\_eff & Rhat &\\ 
\hline
\endhead
\hline
\endfoot
\hline
\endlastfoot
intercept & -1.47 & 0.01 & 0.45 & -1.95 & -1.54 & -0.88 & 4417.49 & 1.00 & 0.01 \\ 
  Sex:Female & 0.00 & 0.00 & 0.08 & -0.10 & 0.00 & 0.11 & 6698.06 & 1.00 & 0.51 \\ 
  sigma Age & 0.12 & 0.00 & 0.14 & 0.02 & 0.08 & 0.27 & 6683.18 & 1.00 & 1.00 \\ 
  Age:-18 & -0.02 & 0.00 & 0.12 & -0.14 & -0.00 & 0.09 & 7073.03 & 1.00 & 0.45 \\ 
  Age:19-29 & 0.03 & 0.00 & 0.11 & -0.06 & 0.02 & 0.15 & 7396.11 & 1.00 & 0.66 \\ 
  Age:30-39 & 0.00 & 0.00 & 0.11 & -0.10 & 0.00 & 0.11 & 6945.41 & 1.00 & 0.54 \\ 
  Age:40+ & -0.05 & 0.00 & 0.11 & -0.17 & -0.03 & 0.03 & 6789.32 & 1.00 & 0.25 \\ 
  sigma State & 0.08 & 0.00 & 0.05 & 0.01 & 0.07 & 0.15 & 6604.02 & 1.00 & 1.00 \\
\end{longtable}

\newpage
\scriptsize
\centering 
\begin{longtable}{rrrrrrrrrr}
\caption{Summary of the posterior distribution for the parameters of the model predicting ``likelihood to spread fake news'' using the model-based ratings of Republican personas - $\nu^{ \{ \mbox{\scriptsize model}, \mbox{ \scriptsize party} \} }_{j=R}$. Fake news is defined as news that is rated at or below \it{Not Very Accurate}.} \\

\hline
coefficient & mean & se\_mean & sd & 10\% & 50\% & 90\% & n\_eff & Rhat & $P(\beta >0)$ \\ 
  \hline
\endfirsthead
\hline
coefficient & mean & se\_mean & sd & 10\% & 50\% & 90\% & n\_eff & Rhat & $P(\beta >0)$ \\ 
\hline
\endhead
\hline
\endfoot
\hline
\endlastfoot
intercept & -1.98 & 0.01 & 0.33 & -2.32 & -2.03 & -1.55 & 2424.67 & 1.00 & 0.00 \\ 
  Sex:Female & -0.10 & 0.00 & 0.09 & -0.21 & -0.09 & 0.02 & 7179.78 & 1.00 & 0.14 \\ 
  sigma Age & 0.24 & 0.00 & 0.20 & 0.08 & 0.19 & 0.48 & 3418.60 & 1.00 & 1.00 \\ 
  Age:-18 & -0.09 & 0.00 & 0.21 & -0.32 & -0.06 & 0.11 & 2280.73 & 1.00 & 0.32 \\ 
  Age:19-29 & -0.08 & 0.00 & 0.19 & -0.27 & -0.06 & 0.09 & 1879.81 & 1.00 & 0.29 \\ 
  Age:30-39 & -0.12 & 0.00 & 0.20 & -0.33 & -0.09 & 0.06 & 1825.33 & 1.00 & 0.22 \\ 
  Age:40+ & 0.11 & 0.00 & 0.18 & -0.05 & 0.11 & 0.29 & 1889.17 & 1.00 & 0.83 \\ 
  sigma State & 0.30 & 0.00 & 0.06 & 0.23 & 0.30 & 0.39 & 6039.33 & 1.00 & 1.00 \\
\end{longtable}

\newpage
\scriptsize
\centering 
\begin{longtable}{rrrrrrrrrr}
\caption{Summary of the posterior distribution for the parameters of the model predicting ``likelihood to spread fake news'' using the model-based ratings of the survey participants -  $\nu^{ \{ \mbox{\scriptsize model}, \mbox{ \scriptsize sample} \} }$. Fake news is defined as news that is rated at or below \it{Not Very Accurate}.} \\

\hline
coefficient & mean & se\_mean & sd & 10\% & 50\% & 90\% & n\_eff & Rhat & $P(\beta >0)$\\ 
  \hline
\endfirsthead
\hline
coefficient & mean & se\_mean & sd & 10\% & 50\% & 90\% & n\_eff & Rhat & $P(\beta >0)$\\ 
\hline
\endhead
\hline
\endfoot
\hline
\endlastfoot
intercept & -1.96 & 0.01 & 1.02 & -3.27 & -1.95 & -0.65 & 7935.59 & 1.00 & 0.03 \\ 
  Sex:Female & -0.46 & 0.01 & 0.88 & -1.60 & -0.44 & 0.66 & 8314.77 & 1.00 & 0.30 \\ 
  sigma Age & 1.41 & 0.02 & 1.10 & 0.21 & 1.11 & 3.15 & 3383.83 & 1.00 & 1.00 \\ 
  Age:-18 & -1.20 & 0.04 & 2.19 & -4.50 & -0.34 & 0.56 & 3007.10 & 1.00 & 0.32 \\ 
  Age:19-29 & -1.60 & 0.04 & 2.39 & -5.32 & -0.63 & 0.33 & 2865.34 & 1.00 & 0.25 \\ 
  Age:30-39 & -1.50 & 0.04 & 2.36 & -5.15 & -0.54 & 0.39 & 2983.58 & 1.00 & 0.27 \\ 
  Age:40+ & -2.02 & 0.05 & 2.60 & -6.15 & -0.99 & 0.12 & 2948.82 & 1.00 & 0.17 \\ 
  sigma State & 0.52 & 0.00 & 0.40 & 0.08 & 0.44 & 1.09 & 7583.39 & 1.00 & 1.00 \\
\end{longtable}

\newpage
\scriptsize
\centering 
\begin{longtable}{rrrrrrrrrr}
\caption{Summary of the posterior distribution for the parameters of the model predicting ``likelihood to spread fake news'' the model-based ratings of Democrat and Republican personas - $\nu^{ \{ \mbox{\scriptsize model}, \mbox{ \scriptsize balanced} \} }$. Fake news is defined as news that is rated at or below \it{Not Very Accurate},}  \\

\hline
coefficient & mean & se\_mean & sd & 10\% & 50\% & 90\% & n\_eff & Rhat & $P(\beta >0)$ \\ 
  \hline
\endfirsthead
\hline
coefficient & mean & se\_mean & sd & 10\% & 50\% & 90\% & n\_eff & Rhat & $P(\beta >0)$ \\ 
\hline
\endhead
\hline
\endfoot
\hline
\endlastfoot
intercept & -2.62 & 0.02 & 0.50 & -3.11 & -2.76 & -1.87 & 715.94 & 1.01 & 0.00 \\ 
  Sex:Female & -0.19 & 0.00 & 0.14 & -0.37 & -0.19 & -0.00 & 3974.34 & 1.00 & 0.09 \\ 
  sigma Age & 0.16 & 0.00 & 0.20 & 0.02 & 0.10 & 0.36 & 1998.63 & 1.00 & 1.00 \\ 
  Age:-18 & -0.06 & 0.00 & 0.18 & -0.26 & -0.02 & 0.08 & 2161.55 & 1.00 & 0.38 \\ 
  Age:19-29 & -0.06 & 0.00 & 0.17 & -0.22 & -0.01 & 0.07 & 1462.16 & 1.01 & 0.38 \\ 
  Age:30-39 & -0.01 & 0.00 & 0.15 & -0.14 & 0.00 & 0.13 & 2460.32 & 1.00 & 0.52 \\ 
  Age:40+ & -0.02 & 0.00 & 0.15 & -0.15 & -0.00 & 0.11 & 1519.51 & 1.01 & 0.49 \\ 
  sigma State & 0.29 & 0.00 & 0.14 & 0.11 & 0.29 & 0.47 & 3715.94 & 1.00 & 1.00 \\
\end{longtable}

\newpage
\scriptsize
\centering 
\begin{longtable}{rrrrrrrrrr}

\caption{Summary of the posterior distribution for the parameters of the model predicting ``likelihood to spread fake news'' using the wisdom of the synthetic crowds - $\nu^{ \{ \mbox{\scriptsize model}, \mbox{ \scriptsize pop.} \} }$. Fake news is defined as news that is rated at or below \it{Not Very Accurate}.} \\

\hline
coefficient & mean & se\_mean & sd & 10\% & 50\% & 90\% & n\_eff & Rhat & $P(\beta >0)$ \\ 
  \hline
\endfirsthead
\hline
coefficient& mean & se\_mean & sd & 10\% & 50\% & 90\% & n\_eff & Rhat & $P(\beta >0)$ \\ 
\hline
\endhead
\hline
\endfoot
\hline
\endlastfoot
intercept & -2.28 & 0.01 & 0.34 & -2.63 & -2.35 & -1.83 & 1853.93 & 1.00 & 0.00 \\ 
  Sex:Female & -0.20 & 0.00 & 0.11 & -0.35 & -0.20 & -0.05 & 7226.83 & 1.00 & 0.04 \\ 
  sigma Age & 0.13 & 0.00 & 0.16 & 0.02 & 0.08 & 0.29 & 5136.71 & 1.00 & 1.00 \\ 
  Age:-18 & -0.04 & 0.00 & 0.14 & -0.20 & -0.01 & 0.08 & 5366.55 & 1.00 & 0.40 \\ 
  Age:19-29 & -0.03 & 0.00 & 0.12 & -0.15 & -0.01 & 0.07 & 5588.90 & 1.00 & 0.40 \\ 
  Age:30-39 & -0.03 & 0.00 & 0.12 & -0.16 & -0.01 & 0.08 & 5707.28 & 1.00 & 0.43 \\ 
  Age:40+ & 0.02 & 0.00 & 0.11 & -0.08 & 0.01 & 0.13 & 5365.16 & 1.00 & 0.61 \\ 
  sigma State & 0.19 & 0.00 & 0.09 & 0.06 & 0.18 & 0.31 & 4806.86 & 1.00 & 1.00 \\
\end{longtable}

\newpage

\section{Fake News Sharing State Estimates\\Using 10\% Threshold}

\begin{table}[] 
\centering
\begin{minipage}[t]{0.45\textwidth}
\centering
\begin{longtable}{rrrrl}
\caption{State-wide estimations of fake news sharing likelihood, using the aggregated ratings of Democratic survey participants - $\nu^{ \{ \mbox{\scriptsize naive}, \mbox{ \scriptsize party} \} }_{j=D}$.}\\
  \hline
  State & 10\% & 50\% & 90\%  \\ 
  \hline
  \endfirsthead
  \hline
  State & 10\% & 50\% & 90\% \\ 
  \hline
  \endhead
  \hline
  \endfoot
  \hline
  \endlastfoot

ND & 0.12 & 0.14 & 0.17 \\ 
  ME & 0.12 & 0.14 & 0.17 \\ 
  WV & 0.11 & 0.14 & 0.15 \\ 
  UT & 0.11 & 0.13 & 0.15 \\ 
  NE & 0.12 & 0.13 & 0.16 \\ 
  TN & 0.11 & 0.13 & 0.15 \\ 
  NH & 0.10 & 0.12 & 0.15 \\ 
  IA & 0.11 & 0.12 & 0.15 \\ 
  KY & 0.11 & 0.12 & 0.14 \\ 
  WI & 0.10 & 0.12 & 0.14 \\ 
  AL & 0.11 & 0.12 & 0.14 \\ 
  ID & 0.10 & 0.12 & 0.14 \\ 
  OH & 0.10 & 0.12 & 0.13 \\ 
  IN & 0.10 & 0.12 & 0.14 \\ 
  RI & 0.09 & 0.12 & 0.15 \\ 
  KS & 0.10 & 0.12 & 0.13 \\ 
  AR & 0.10 & 0.12 & 0.13 \\ 
  MI & 0.10 & 0.11 & 0.14 \\ 
  PA & 0.10 & 0.11 & 0.13 \\ 
  MO & 0.10 & 0.11 & 0.13 \\ 
  MT & 0.09 & 0.11 & 0.13 \\ 
  MN & 0.09 & 0.11 & 0.13 \\ 
  OK & 0.09 & 0.11 & 0.13 \\ 
  SC & 0.09 & 0.11 & 0.12 \\ 
  SD & 0.09 & 0.11 & 0.13 \\ 
  VT & 0.09 & 0.11 & 0.14 \\ 
  CO & 0.10 & 0.11 & 0.13 \\ 
  LA & 0.09 & 0.11 & 0.13 \\ 
  AZ & 0.09 & 0.11 & 0.12 \\ 
  NC & 0.09 & 0.11 & 0.12 \\ 
  WY & 0.09 & 0.11 & 0.13 \\ 
  TX & 0.09 & 0.11 & 0.12 \\ 
  FL & 0.09 & 0.11 & 0.12 \\ 
  MS & 0.09 & 0.11 & 0.12 \\ 
  VA & 0.09 & 0.10 & 0.12 \\ 
  AK & 0.08 & 0.10 & 0.13 \\ 
  IL & 0.09 & 0.10 & 0.12 \\ 
  NY & 0.09 & 0.10 & 0.12 \\ 
  CT & 0.09 & 0.10 & 0.12 \\ 
  GA & 0.09 & 0.10 & 0.12 \\ 
  OR & 0.08 & 0.10 & 0.11 \\ 
  NJ & 0.08 & 0.09 & 0.12 \\ 
  DE & 0.08 & 0.09 & 0.11 \\ 
  MA & 0.08 & 0.09 & 0.11 \\ 
  WA & 0.08 & 0.09 & 0.10 \\ 
  MD & 0.07 & 0.09 & 0.10 \\ 
  NV & 0.07 & 0.09 & 0.10 \\ 
  NM & 0.07 & 0.08 & 0.10 \\ 
  CA & 0.07 & 0.08 & 0.09 \\ 
  HI & 0.05 & 0.07 & 0.09 \\ 
  DC & 0.02 & 0.04 & 0.06 \\ 
  \hline
\end{longtable}
\end{minipage}%
\hspace{0.05\textwidth}
\begin{minipage}[t]{0.45\textwidth}
\centering
\vspace{.5cm}
\begin{longtable}{lrrr}
\caption{Predicted amount of fake news sharers per state, using the aggregated ratings of Democratic survey participants - $\nu^{ \{ \mbox{\scriptsize naive}, \mbox{ \scriptsize party} \} }_{j=D}$.} \\
  \hline
  State & 10\% & 50\% & 90\% \\ 
  \hline
  \endfirsthead
  \hline
  State & 10\% & 50\% & 90\% \\ 
  \hline
  \endhead
  \hline
  \endfoot
  \hline
  \endlastfoot
CA & 26708 & 31228 & 36353 \\ 
  TX & 24307 & 27834 & 32061 \\ 
  FL & 18977 & 21455 & 23893 \\ 
  NY & 18073 & 20415 & 23684 \\ 
  PA & 12675 & 14713 & 16620 \\ 
  OH & 12518 & 14338 & 16166 \\ 
  IL & 11537 & 13036 & 15149 \\ 
  NC & 9862 & 11383 & 12972 \\ 
  MI & 10102 & 11381 & 13456 \\ 
  GA & 8550 & 9876 & 11368 \\ 
  VA & 7850 & 9082 & 10470 \\ 
  TN & 7910 & 8944 & 10213 \\ 
  NJ & 7505 & 8768 & 11153 \\ 
  IN & 7059 & 8068 & 9240 \\ 
  AZ & 6733 & 7984 & 9098 \\ 
  WI & 6322 & 7324 & 8490 \\ 
  WA & 6237 & 7173 & 8078 \\ 
  MO & 6028 & 7016 & 7936 \\ 
  MA & 5489 & 6766 & 8207 \\ 
  CO & 5634 & 6380 & 7348 \\ 
  MN & 5378 & 6325 & 7292 \\ 
  AL & 5313 & 5994 & 7151 \\ 
  SC & 4828 & 5691 & 6428 \\ 
  KY & 4942 & 5603 & 6519 \\ 
  MD & 4591 & 5415 & 6314 \\ 
  LA & 4046 & 4791 & 5554 \\ 
  UT & 3810 & 4473 & 5192 \\ 
  OK & 3458 & 4181 & 4925 \\ 
  OR & 3551 & 4130 & 4649 \\ 
  IA & 3489 & 4024 & 4729 \\ 
  CT & 3233 & 3731 & 4380 \\ 
  AR & 2993 & 3467 & 4044 \\ 
  KS & 2842 & 3386 & 3874 \\ 
  MS & 2566 & 3008 & 3528 \\ 
  NE & 2273 & 2576 & 3158 \\ 
  NV & 2094 & 2569 & 3009 \\ 
  WV & 1992 & 2355 & 2701 \\ 
  ID & 1817 & 2156 & 2571 \\ 
  ME & 1633 & 1955 & 2368 \\ 
  NH & 1397 & 1695 & 2017 \\ 
  NM & 1307 & 1589 & 1904 \\ 
  RI & 1035 & 1296 & 1644 \\ 
  MT & 1007 & 1191 & 1375 \\ 
  ND & 952 & 1113 & 1283 \\ 
  HI & 793 & 1040 & 1384 \\ 
  SD & 822 & 973 & 1142 \\ 
  DE & 747 & 882 & 1026 \\ 
  VT & 586 & 734 & 922 \\ 
  AK & 524 & 663 & 829 \\ 
  WY & 528 & 635 & 742 \\ 
  DC & 129 & 254 & 442 \\ 
  \hline
\end{longtable}
\end{minipage}
\label{statedist_naivedem}
\end{table}

\begin{table}[] 
\centering
\begin{minipage}[t]{0.45\textwidth}
\centering
\begin{longtable}{rrrrl}
\caption{Summary of the posterior distribution for the parameters of the model predicting ``likelihood to spread fake news'', using the aggregated ratings of Republican survey participants - $\nu^{ \{ \mbox{\scriptsize naive}, \mbox{ \scriptsize party} \} }_{j=R}$.} \\
  \hline
  State & 10\% & 50\% & 90\%  \\ 
  \hline
  \endfirsthead
  \hline
  State & 10\% & 50\% & 90\% \\ 
  \hline
  \endhead
  \hline
  \endfoot
  \hline
  \endlastfoot
HI & 0.09 & 0.12 & 0.16 \\ 
  NM & 0.09 & 0.12 & 0.14 \\ 
  MI & 0.10 & 0.12 & 0.15 \\ 
  CA & 0.10 & 0.12 & 0.13 \\ 
  NV & 0.10 & 0.12 & 0.14 \\ 
  MD & 0.10 & 0.12 & 0.14 \\ 
  TX & 0.10 & 0.12 & 0.13 \\ 
  IN & 0.10 & 0.11 & 0.15 \\
\hline
\end{longtable}
\end{minipage}%
\hspace{0.05\textwidth}
\begin{minipage}[t]{0.45\textwidth}
\centering

\begin{longtable}{lrrr}
\caption{Predicted amount of fake news sharers per state, using the aggregated ratings of Republican survey participants - $\nu^{ \{ \mbox{\scriptsize naive}, \mbox{ \scriptsize party} \} }_{j=R}$.} \\
  \hline
  State & 10\% & 50\% & 90\% \\ 
  \hline
  \endfirsthead
  \hline
  State & 10\% & 50\% & 90\% \\ 
  \hline
  \endhead
  \hline
  \endfoot
  \hline
  \endlastfoot
CA & 39346 & 45301 & 51956 \\ 
  TX & 26373 & 30237 & 34819 \\ 
  NY & 19759 & 22378 & 25520 \\ 
  FL & 19604 & 21958 & 24739 \\ 
  IL & 10992 & 13524 & 15278 \\ 
  PA & 11274 & 13198 & 14999 \\ 
  OH & 10285 & 12137 & 13806 \\ 
  MI & 10162 & 11629 & 14938 \\ 
  NC & 9050 & 10847 & 12383 \\ 
  NJ & 8761 & 10478 & 12108 \\ 
  GA & 8178 & 10187 & 11666 \\ 
  VA & 7417 & 8969 & 10315 \\ 
  WA & 7377 & 8858 & 9976 \\ 
  AZ & 6856 & 7912 & 9195 \\ 
  IN & 6792 & 7817 & 10152 \\ 
  MA & 6282 & 7697 & 9303 \\ 
  MD & 6256 & 7302 & 8624 \\ 
  TN & 5851 & 6956 & 7952 \\ 
  MO & 5623 & 6624 & 7616 \\ 
  WI & 5193 & 6262 & 7312 \\ 
  CO & 4969 & 6071 & 6873 \\ 
  MN & 4831 & 5969 & 6943 \\ 
  SC & 4384 & 5335 & 6073 \\ 
  AL & 4634 & 5300 & 6385 \\ 
  OR & 4253 & 4892 & 5631 \\ 
  KY & 3906 & 4705 & 5410 \\ 
  LA & 3957 & 4702 & 5541 \\ 
  OK & 3516 & 4185 & 5017 \\ 
  CT & 3033 & 3798 & 4377 \\ 
  NV & 2904 & 3521 & 4124 \\ 
  IA & 2760 & 3348 & 3934 \\ 
  AR & 2792 & 3290 & 3908 \\ 
  UT & 2580 & 3230 & 3765 \\ 
  KS & 2706 & 3143 & 3788 \\ 
  MS & 2518 & 3081 & 3603 \\ 
  NM & 1825 & 2299 & 2736 \\ 
  NE & 1642 & 1979 & 2310 \\ 
  ID & 1541 & 1903 & 2255 \\ 
  HI & 1334 & 1816 & 2348 \\ 
  WV & 1411 & 1707 & 2008 \\ 
  NH & 1157 & 1408 & 1738 \\ 
  ME & 1086 & 1332 & 1625 \\ 
  MT & 933 & 1148 & 1335 \\ 
  RI & 815 & 1051 & 1324 \\ 
  DE & 840 & 1028 & 1190 \\ 
  SD & 801 & 980 & 1155 \\ 
  DC & 480 & 744 & 1071 \\ 
  ND & 605 & 735 & 877 \\ 
  AK & 556 & 725 & 897 \\ 
  VT & 555 & 703 & 896 \\ 
  WY & 533 & 652 & 776 \\ 
  \hline
\end{longtable}
\end{minipage}
\label{statedist_naiverep}
\end{table}

\begin{table}[] 
\centering
\begin{minipage}[t]{0.45\textwidth}
\centering
\begin{longtable}{rrrrl}
\caption{State-wide estimations of fake news sharing likelihood, using the average survey rating - $\nu^{ \{ \mbox{\scriptsize naive}, \mbox{ \scriptsize sample} \} }$.} \\
  \hline
  State & 10\% & 50\% & 90\%  \\ 
  \hline
  \endfirsthead
  \hline
  State & 10\% & 50\% & 90\% \\ 
  \hline
  \endhead
  \hline
  \endfoot
  \hline
  \endlastfoot

NE & 0.13 & 0.15 & 0.19 \\ 
  IN & 0.13 & 0.15 & 0.18 \\ 
  ND & 0.12 & 0.14 & 0.17 \\ 
  WY & 0.12 & 0.14 & 0.17 \\ 
  AR & 0.12 & 0.14 & 0.17 \\ 
  ME & 0.12 & 0.14 & 0.18 \\ 
  ID & 0.12 & 0.14 & 0.17 \\ 
  KY & 0.12 & 0.14 & 0.16 \\ 
  WI & 0.12 & 0.14 & 0.16 \\ 
  WV & 0.11 & 0.14 & 0.16 \\ 
  IA & 0.12 & 0.14 & 0.16 \\ 
  MI & 0.12 & 0.14 & 0.16 \\ 
  FL & 0.12 & 0.14 & 0.16 \\ 
  UT & 0.11 & 0.14 & 0.16 \\ 
  NJ & 0.12 & 0.14 & 0.17 \\ 
  KS & 0.11 & 0.14 & 0.16 \\ 
  MT & 0.11 & 0.14 & 0.16 \\ 
  MO & 0.12 & 0.14 & 0.16 \\ 
  OR & 0.12 & 0.14 & 0.16 \\ 
  NH & 0.11 & 0.14 & 0.17 \\ 
  TX & 0.12 & 0.14 & 0.15 \\ 
  TN & 0.12 & 0.14 & 0.15 \\ 
  SD & 0.11 & 0.13 & 0.16 \\ 
  CO & 0.12 & 0.13 & 0.16 \\ 
  MN & 0.11 & 0.13 & 0.15 \\ 
  AK & 0.10 & 0.13 & 0.16 \\ 
  VT & 0.11 & 0.13 & 0.17 \\ 
  OK & 0.11 & 0.13 & 0.15 \\ 
  AL & 0.11 & 0.13 & 0.15 \\ 
  WA & 0.11 & 0.13 & 0.15 \\ 
  MS & 0.11 & 0.13 & 0.15 \\ 
  IL & 0.11 & 0.13 & 0.14 \\ 
  NV & 0.11 & 0.13 & 0.15 \\ 
  PA & 0.11 & 0.13 & 0.14 \\ 
  OH & 0.10 & 0.13 & 0.14 \\ 
  LA & 0.10 & 0.13 & 0.15 \\ 
  NC & 0.11 & 0.13 & 0.14 \\ 
  CT & 0.11 & 0.12 & 0.15 \\ 
  SC & 0.10 & 0.12 & 0.14 \\ 
  NY & 0.11 & 0.12 & 0.14 \\ 
  DE & 0.10 & 0.12 & 0.15 \\ 
  AZ & 0.10 & 0.12 & 0.14 \\ 
  GA & 0.10 & 0.12 & 0.14 \\ 
  VA & 0.10 & 0.12 & 0.14 \\ 
  MD & 0.10 & 0.12 & 0.14 \\ 
  RI & 0.09 & 0.12 & 0.15 \\ 
  NM & 0.09 & 0.12 & 0.14 \\ 
  MA & 0.09 & 0.11 & 0.14 \\ 
  CA & 0.10 & 0.11 & 0.13 \\ 
  HI & 0.08 & 0.11 & 0.14 \\ 
  DC & 0.05 & 0.08 & 0.12 \\ 
\hline
\end{longtable}
\end{minipage}%
\hspace{0.05\textwidth}
\begin{minipage}[t]{0.45\textwidth}
\centering

\begin{longtable}{lrrr}
\caption{Predicted amount of fake news sharers per state, using the average survey rating - $\nu^{ \{ \mbox{\scriptsize naive}, \mbox{ \scriptsize sample} \} }$.} \\
  \hline
  State & 10\% & 50\% & 90\% \\ 
  \hline
  \endfirsthead
  \hline
  State & 10\% & 50\% & 90\% \\ 
  \hline
  \endhead
  \hline
  \endfoot
  \hline
  \endlastfoot
CA & 37697 & 43087 & 49121 \\ 
  TX & 31238 & 35440 & 40206 \\ 
  FL & 25118 & 27920 & 31819 \\ 
  NY & 21961 & 24688 & 27851 \\ 
  PA & 13719 & 16326 & 18495 \\ 
  IL & 13643 & 16065 & 18128 \\ 
  OH & 12330 & 15077 & 17202 \\ 
  MI & 12021 & 13686 & 15872 \\ 
  NC & 10962 & 13089 & 15017 \\ 
  NJ & 11003 & 12727 & 15843 \\ 
  GA & 9480 & 11622 & 13367 \\ 
  VA & 8590 & 10324 & 11854 \\ 
  WA & 8899 & 10252 & 11749 \\ 
  IN & 8755 & 9958 & 11992 \\ 
  TN & 8046 & 9466 & 10811 \\ 
  AZ & 6947 & 8757 & 10116 \\ 
  MO & 7256 & 8488 & 9728 \\ 
  WI & 7044 & 8403 & 9742 \\ 
  MA & 6907 & 8398 & 10155 \\ 
  CO & 6875 & 7819 & 9151 \\ 
  MN & 6114 & 7548 & 8740 \\ 
  MD & 6202 & 7397 & 8656 \\ 
  AL & 5477 & 6493 & 7476 \\ 
  KY & 5447 & 6407 & 7435 \\ 
  SC & 5165 & 6383 & 7300 \\ 
  OR & 5129 & 5818 & 6888 \\ 
  LA & 4363 & 5478 & 6341 \\ 
  OK & 4024 & 4957 & 5800 \\ 
  UT & 3850 & 4610 & 5452 \\ 
  CT & 3853 & 4546 & 5375 \\ 
  IA & 3775 & 4514 & 5314 \\ 
  AR & 3655 & 4264 & 5096 \\ 
  KS & 3249 & 3952 & 4562 \\ 
  NV & 3186 & 3791 & 4392 \\ 
  MS & 3076 & 3683 & 4398 \\ 
  NE & 2512 & 2884 & 3675 \\ 
  ID & 2065 & 2513 & 3005 \\ 
  WV & 1985 & 2424 & 2825 \\ 
  NM & 1755 & 2288 & 2737 \\ 
  ME & 1641 & 1968 & 2461 \\ 
  NH & 1532 & 1847 & 2249 \\ 
  HI & 1175 & 1596 & 2062 \\ 
  MT & 1187 & 1448 & 1675 \\ 
  RI & 999 & 1284 & 1615 \\ 
  SD & 955 & 1190 & 1386 \\ 
  DE & 970 & 1157 & 1382 \\ 
  ND & 909 & 1098 & 1311 \\ 
  VT & 705 & 880 & 1121 \\ 
  AK & 665 & 852 & 1046 \\ 
  WY & 704 & 837 & 1011 \\ 
  DC & 368 & 573 & 849 \\ 
  \hline
\end{longtable}
\end{minipage}
\label{statedist_naivesample}
\end{table}

\begin{table}[] 
\centering
\begin{minipage}[t]{0.45\textwidth}
\centering
\begin{longtable}{rrrrl}
\caption{State-wide estimations of fake news sharing likelihood, using as many Republicans as Democrats per tweets - $\nu^{ \{ \mbox{\scriptsize naive}, \mbox{ \scriptsize balanced} \} }$.} \\
  \hline
  State & 10\% & 50\% & 90\%  \\ 
  \hline
  \endfirsthead
  \hline
  State & 10\% & 50\% & 90\% \\ 
  \hline
  \endhead
  \hline
  \endfoot
  \hline
  \endlastfoot
ME & 0.10 & 0.12 & 0.14 \\ 
  ND & 0.10 & 0.11 & 0.13 \\ 
  NH & 0.10 & 0.11 & 0.14 \\ 
  UT & 0.10 & 0.11 & 0.13 \\ 
  WV & 0.10 & 0.11 & 0.13 \\ 
  IA & 0.10 & 0.11 & 0.13 \\ 
  NE & 0.10 & 0.11 & 0.13 \\ 
  VT & 0.09 & 0.11 & 0.14 \\ 
  WI & 0.10 & 0.11 & 0.13 \\ 
  IN & 0.10 & 0.11 & 0.13 \\ 
  KY & 0.10 & 0.11 & 0.12 \\ 
  TN & 0.10 & 0.11 & 0.12 \\ 
  KS & 0.10 & 0.11 & 0.12 \\ 
  ID & 0.09 & 0.11 & 0.13 \\ 
  MN & 0.09 & 0.11 & 0.12 \\ 
  MI & 0.10 & 0.11 & 0.12 \\ 
  PA & 0.09 & 0.11 & 0.12 \\ 
  MT & 0.09 & 0.11 & 0.12 \\ 
  RI & 0.09 & 0.11 & 0.13 \\ 
  OH & 0.09 & 0.11 & 0.12 \\ 
  SD & 0.09 & 0.11 & 0.12 \\ 
  AL & 0.09 & 0.11 & 0.12 \\ 
  WY & 0.09 & 0.11 & 0.12 \\ 
  CO & 0.09 & 0.11 & 0.12 \\ 
  MO & 0.09 & 0.11 & 0.12 \\ 
  AR & 0.09 & 0.11 & 0.12 \\ 
  OR & 0.09 & 0.10 & 0.12 \\ 
  SC & 0.09 & 0.10 & 0.12 \\ 
  IL & 0.09 & 0.10 & 0.11 \\ 
  NC & 0.09 & 0.10 & 0.11 \\ 
  MA & 0.08 & 0.10 & 0.12 \\ 
  VA & 0.09 & 0.10 & 0.12 \\ 
  CT & 0.09 & 0.10 & 0.12 \\ 
  TX & 0.09 & 0.10 & 0.12 \\ 
  OK & 0.08 & 0.10 & 0.12 \\ 
  MS & 0.09 & 0.10 & 0.12 \\ 
  FL & 0.09 & 0.10 & 0.11 \\ 
  GA & 0.09 & 0.10 & 0.11 \\ 
  AZ & 0.09 & 0.10 & 0.11 \\ 
  LA & 0.09 & 0.10 & 0.11 \\ 
  DE & 0.09 & 0.10 & 0.11 \\ 
  NY & 0.09 & 0.10 & 0.11 \\ 
  WA & 0.09 & 0.10 & 0.11 \\ 
  AK & 0.08 & 0.10 & 0.12 \\ 
  MD & 0.08 & 0.10 & 0.11 \\ 
  NJ & 0.08 & 0.10 & 0.11 \\ 
  NV & 0.08 & 0.09 & 0.11 \\ 
  CA & 0.08 & 0.09 & 0.11 \\ 
  NM & 0.08 & 0.09 & 0.11 \\ 
  HI & 0.06 & 0.08 & 0.11 \\ 
  DC & 0.04 & 0.06 & 0.10 \\ 
\hline
\end{longtable}
\end{minipage}%
\hspace{0.05\textwidth}
\begin{minipage}[t]{0.45\textwidth}
\centering

\begin{longtable}{lrrr}
\caption{Predicted amount of fake news sharers per state. Using as many Republicans as Democrats per tweets - $\nu^{ \{ \mbox{\scriptsize naive}, \mbox{ \scriptsize balanced} \} }$.} \\
  \hline
  State & 10\% & 50\% & 90\% \\ 
  \hline
  \endfirsthead
  \hline
  State & 10\% & 50\% & 90\% \\ 
  \hline
  \endhead
  \hline
  \endfoot
  \hline
  \endlastfoot
CA & 31207 & 36379 & 42398 \\ 
  TX & 23080 & 26463 & 30244 \\ 
  FL & 18433 & 20470 & 22743 \\ 
  NY & 17794 & 20042 & 22579 \\ 
  PA & 12217 & 13882 & 15501 \\ 
  IL & 11651 & 12976 & 14426 \\ 
  OH & 11145 & 12735 & 14182 \\ 
  NC & 9375 & 10691 & 11944 \\ 
  MI & 9563 & 10642 & 11988 \\ 
  GA & 8666 & 9797 & 11097 \\ 
  NJ & 7727 & 8870 & 10147 \\ 
  VA & 7700 & 8850 & 10059 \\ 
  WA & 6946 & 7844 & 8719 \\ 
  TN & 6816 & 7611 & 8576 \\ 
  IN & 6729 & 7517 & 8574 \\ 
  MA & 6155 & 7439 & 8877 \\ 
  AZ & 6361 & 7357 & 8261 \\ 
  WI & 5847 & 6654 & 7625 \\ 
  MO & 5696 & 6549 & 7297 \\ 
  CO & 5499 & 6124 & 6845 \\ 
  MN & 5315 & 6107 & 7000 \\ 
  MD & 5192 & 6008 & 6850 \\ 
  SC & 4724 & 5332 & 5969 \\ 
  AL & 4663 & 5243 & 5873 \\ 
  KY & 4437 & 5035 & 5698 \\ 
  OR & 4011 & 4456 & 4975 \\ 
  LA & 3774 & 4395 & 4978 \\ 
  OK & 3199 & 3816 & 4444 \\ 
  UT & 3300 & 3785 & 4331 \\ 
  CT & 3282 & 3724 & 4227 \\ 
  IA & 3200 & 3646 & 4180 \\ 
  AR & 2764 & 3160 & 3609 \\ 
  KS & 2769 & 3149 & 3558 \\ 
  MS & 2514 & 2885 & 3315 \\ 
  NV & 2413 & 2849 & 3292 \\ 
  NE & 1923 & 2173 & 2482 \\ 
  WV & 1709 & 1954 & 2236 \\ 
  ID & 1665 & 1941 & 2250 \\ 
  NM & 1545 & 1832 & 2155 \\ 
  ME & 1356 & 1601 & 1906 \\ 
  NH & 1305 & 1538 & 1840 \\ 
  HI & 963 & 1258 & 1613 \\ 
  RI & 939 & 1168 & 1455 \\ 
  MT & 993 & 1134 & 1285 \\ 
  DE & 832 & 945 & 1074 \\ 
  SD & 812 & 935 & 1071 \\ 
  ND & 767 & 878 & 1010 \\ 
  VT & 604 & 745 & 920 \\ 
  WY & 531 & 619 & 714 \\ 
  AK & 493 & 618 & 757 \\ 
  DC & 259 & 452 & 722 \\ 
  \hline
\end{longtable}
\end{minipage}
\label{statedist_naivebal}
\end{table}

\begin{table}[] 
\centering
\begin{minipage}[t]{0.45\textwidth}
\centering
\begin{longtable}{rrrrl}
\caption{State-wide estimations of fake news sharing likelihood. Using the weighted assessment of the survey participants - $\nu^{ \{ \mbox{\scriptsize naive}, \mbox{ \scriptsize pop.} \} }.$}  \\

  \hline
  State & 10\% & 50\% & 90\%  \\ 
  \hline
  \endfirsthead
  \hline
  State & 10\% & 50\% & 90\% \\ 
  \hline
  \endhead
  \hline
  \endfoot
  \hline
  \endlastfoot

NE & 0.10 & 0.12 & 0.17 \\ 
  ID & 0.09 & 0.12 & 0.15 \\ 
  AK & 0.09 & 0.12 & 0.15 \\ 
  IN & 0.10 & 0.12 & 0.15 \\ 
  AR & 0.10 & 0.12 & 0.15 \\ 
  NJ & 0.10 & 0.12 & 0.15 \\ 
  WY & 0.09 & 0.11 & 0.14 \\ 
  OK & 0.09 & 0.11 & 0.14 \\ 
  TX & 0.10 & 0.11 & 0.13 \\ 
  MI & 0.09 & 0.11 & 0.14 \\ 
  WI & 0.09 & 0.11 & 0.13 \\ 
  MT & 0.09 & 0.11 & 0.13 \\ 
  KS & 0.09 & 0.11 & 0.13 \\ 
  ND & 0.08 & 0.11 & 0.13 \\ 
  SD & 0.08 & 0.11 & 0.13 \\ 
  MO & 0.09 & 0.11 & 0.13 \\ 
  FL & 0.09 & 0.11 & 0.12 \\ 
  TN & 0.09 & 0.11 & 0.13 \\ 
  MN & 0.08 & 0.10 & 0.13 \\ 
  WV & 0.08 & 0.10 & 0.13 \\ 
  KY & 0.08 & 0.10 & 0.12 \\ 
  OR & 0.09 & 0.10 & 0.13 \\ 
  NV & 0.08 & 0.10 & 0.12 \\ 
  IA & 0.08 & 0.10 & 0.13 \\ 
  LA & 0.08 & 0.10 & 0.12 \\ 
  MS & 0.08 & 0.10 & 0.12 \\ 
  WA & 0.09 & 0.10 & 0.12 \\ 
  CO & 0.09 & 0.10 & 0.12 \\ 
  AL & 0.08 & 0.10 & 0.12 \\ 
  ME & 0.08 & 0.10 & 0.13 \\ 
  IL & 0.08 & 0.10 & 0.11 \\ 
  UT & 0.08 & 0.10 & 0.12 \\ 
  NM & 0.07 & 0.10 & 0.12 \\ 
  AZ & 0.07 & 0.10 & 0.11 \\ 
  NC & 0.07 & 0.09 & 0.11 \\ 
  NH & 0.07 & 0.09 & 0.12 \\ 
  GA & 0.07 & 0.09 & 0.11 \\ 
  NY & 0.08 & 0.09 & 0.11 \\ 
  SC & 0.07 & 0.09 & 0.11 \\ 
  PA & 0.07 & 0.09 & 0.11 \\ 
  DE & 0.07 & 0.09 & 0.11 \\ 
  CT & 0.07 & 0.09 & 0.11 \\ 
  HI & 0.06 & 0.09 & 0.12 \\ 
  OH & 0.07 & 0.09 & 0.11 \\ 
  VA & 0.07 & 0.09 & 0.11 \\ 
  MD & 0.07 & 0.08 & 0.10 \\ 
  CA & 0.07 & 0.08 & 0.10 \\ 
  VT & 0.06 & 0.08 & 0.11 \\ 
  MA & 0.06 & 0.08 & 0.10 \\ 
  RI & 0.06 & 0.08 & 0.10 \\ 
  DC & 0.04 & 0.06 & 0.10 \\ 
\hline
\end{longtable}
\end{minipage}%
\hspace{0.05\textwidth}
\begin{minipage}[t]{0.45\textwidth}
\centering

\begin{longtable}{lrrr}
\caption{Predicted amount of fake news sharers per state. Using the weighted assessment of the survey participants - $\nu^{ \{ \mbox{\scriptsize naive}, \mbox{ \scriptsize pop.} \} }.$}  \\
  \hline
  State & 10\% & 50\% & 90\% \\ 
  \hline
  \endfirsthead
  \hline
  State & 10\% & 50\% & 90\% \\ 
  \hline
  \endhead
  \hline
  \endfoot
  \hline
  \endlastfoot
CA & 27645 & 32532 & 37829 \\ 
  TX & 25221 & 29077 & 33748 \\ 
  FL & 18877 & 21485 & 24829 \\ 
  NY & 16012 & 18505 & 21422 \\ 
  IL & 10040 & 12292 & 14316 \\ 
  PA & 9589 & 11887 & 13795 \\ 
  NJ & 8973 & 10819 & 14291 \\ 
  MI & 9310 & 10795 & 13395 \\ 
  OH & 7913 & 10642 & 12794 \\ 
  NC & 7761 & 9745 & 11583 \\ 
  GA & 6996 & 9006 & 10606 \\ 
  WA & 6763 & 7959 & 9392 \\ 
  IN & 6735 & 7943 & 10019 \\ 
  VA & 6246 & 7662 & 9150 \\ 
  TN & 6041 & 7342 & 8732 \\ 
  AZ & 5388 & 6954 & 8214 \\ 
  MO & 5480 & 6595 & 7805 \\ 
  WI & 5216 & 6541 & 7790 \\ 
  MN & 4616 & 5922 & 7087 \\ 
  CO & 4976 & 5867 & 7131 \\ 
  MA & 4442 & 5660 & 7145 \\ 
  MD & 4098 & 5320 & 6399 \\ 
  AL & 4095 & 4982 & 5961 \\ 
  KY & 3828 & 4758 & 5618 \\ 
  SC & 3648 & 4741 & 5604 \\ 
  LA & 3419 & 4454 & 5412 \\ 
  OR & 3778 & 4422 & 5384 \\ 
  OK & 3339 & 4230 & 5121 \\ 
  AR & 2893 & 3501 & 4453 \\ 
  IA & 2701 & 3358 & 4200 \\ 
  CT & 2661 & 3289 & 4039 \\ 
  UT & 2558 & 3274 & 3998 \\ 
  KS & 2495 & 3126 & 3781 \\ 
  NV & 2501 & 3109 & 3705 \\ 
  MS & 2300 & 2907 & 3564 \\ 
  NE & 1950 & 2315 & 3257 \\ 
  ID & 1696 & 2114 & 2624 \\ 
  NM & 1423 & 1865 & 2324 \\ 
  WV & 1402 & 1821 & 2210 \\ 
  ME & 1075 & 1361 & 1783 \\ 
  HI & 925 & 1331 & 1820 \\ 
  NH & 984 & 1266 & 1630 \\ 
  MT & 908 & 1147 & 1399 \\ 
  SD & 719 & 944 & 1139 \\ 
  DE & 693 & 854 & 1068 \\ 
  RI & 606 & 838 & 1115 \\ 
  ND & 652 & 829 & 1036 \\ 
  AK & 557 & 754 & 963 \\ 
  WY & 520 & 662 & 815 \\ 
  VT & 425 & 558 & 760 \\ 
  DC & 251 & 431 & 688 \\ 
  \hline
\end{longtable}
\end{minipage}
\label{statedist_naivepop}
\end{table}

\begin{table}[] 
\centering
\begin{minipage}[t]{0.45\textwidth}
\centering
\begin{longtable}{rrrrl}
\caption{State-wide estimations of fake news sharing likelihood. Using the model-based ratings of Democrat personas - $\nu^{ \{ \mbox{\scriptsize model}, \mbox{ \scriptsize party} \} }_{j=D}$.}  \\
\\
  \hline
  State & 10\% & 50\% & 90\%  \\ 
  \hline
  \endfirsthead
  \hline
  State & 10\% & 50\% & 90\% \\ 
  \hline
  \endhead
  \hline
  \endfoot
  \hline
  \endlastfoot

AK & 0.10 & 0.13 & 0.15 \\ 
  OK & 0.11 & 0.12 & 0.14 \\ 
  ND & 0.11 & 0.12 & 0.14 \\ 
  ID & 0.10 & 0.12 & 0.14 \\ 
  TX & 0.11 & 0.12 & 0.13 \\ 
  LA & 0.11 & 0.12 & 0.13 \\ 
  AR & 0.10 & 0.12 & 0.13 \\ 
  WV & 0.10 & 0.12 & 0.13 \\ 
  TN & 0.10 & 0.12 & 0.13 \\ 
  NE & 0.10 & 0.11 & 0.13 \\ 
  KS & 0.10 & 0.11 & 0.13 \\ 
  AL & 0.10 & 0.11 & 0.13 \\ 
  AZ & 0.10 & 0.11 & 0.12 \\ 
  MS & 0.10 & 0.11 & 0.13 \\ 
  SD & 0.10 & 0.11 & 0.13 \\ 
  WY & 0.10 & 0.11 & 0.13 \\ 
  KY & 0.10 & 0.11 & 0.12 \\ 
  UT & 0.10 & 0.11 & 0.12 \\ 
  MN & 0.10 & 0.11 & 0.13 \\ 
  WI & 0.10 & 0.11 & 0.12 \\ 
  MT & 0.10 & 0.11 & 0.12 \\ 
  IN & 0.10 & 0.11 & 0.12 \\ 
  SC & 0.10 & 0.11 & 0.12 \\ 
  FL & 0.10 & 0.11 & 0.12 \\ 
  MO & 0.10 & 0.11 & 0.12 \\ 
  OH & 0.10 & 0.11 & 0.12 \\ 
  NV & 0.09 & 0.11 & 0.12 \\ 
  GA & 0.10 & 0.11 & 0.12 \\ 
  IA & 0.09 & 0.11 & 0.12 \\ 
  CO & 0.09 & 0.10 & 0.12 \\ 
  NC & 0.09 & 0.10 & 0.12 \\ 
  ME & 0.09 & 0.10 & 0.12 \\ 
  MI & 0.09 & 0.10 & 0.11 \\ 
  NM & 0.09 & 0.10 & 0.12 \\ 
  PA & 0.09 & 0.10 & 0.11 \\ 
  IL & 0.09 & 0.10 & 0.11 \\ 
  OR & 0.09 & 0.10 & 0.11 \\ 
  HI & 0.08 & 0.10 & 0.12 \\ 
  NH & 0.08 & 0.10 & 0.12 \\ 
  VA & 0.09 & 0.10 & 0.11 \\ 
  WA & 0.09 & 0.10 & 0.11 \\ 
  CA & 0.08 & 0.09 & 0.11 \\ 
  NJ & 0.08 & 0.09 & 0.11 \\ 
  MD & 0.08 & 0.09 & 0.11 \\ 
  NY & 0.08 & 0.09 & 0.10 \\ 
  CT & 0.08 & 0.09 & 0.10 \\ 
  DE & 0.08 & 0.09 & 0.10 \\ 
  RI & 0.07 & 0.09 & 0.11 \\ 
  VT & 0.07 & 0.08 & 0.10 \\ 
  MA & 0.07 & 0.08 & 0.10 \\ 
  DC & 0.02 & 0.03 & 0.06 \\ 
\hline
\end{longtable}%
\end{minipage}
\label{statedist_modelpop}
\end{table}

\newpage
\section{Fake News Sharing State Estimates\\Using the \textit{Not Very Accurate} Threshold}

\begin{table}[] 
\centering
\begin{minipage}[t]{0.45\textwidth}
\centering
\begin{longtable}{rrrrl}
\caption{State-wide estimations of fake news sharing likelihood, using the aggregated ratings of Democratic survey participants - $\nu^{ \{ \mbox{\scriptsize naive}, \mbox{ \scriptsize party} \} }_{j=D}$. Fake news is defined as news that is rated at or below \it{Not Very Accurate}.}\\
  \hline
  State & 10\% & 50\% & 90\%  \\ 
  \hline
  \endfirsthead
  \hline
  State & 10\% & 50\% & 90\% \\ 
  \hline
  \endhead
  \hline
  \endfoot
  \hline
  \endlastfoot
ND & 0.32 & 0.34 & 0.36 \\ 
  WV & 0.31 & 0.33 & 0.35 \\ 
  ME & 0.30 & 0.33 & 0.35 \\ 
  UT & 0.31 & 0.33 & 0.35 \\ 
  TN & 0.30 & 0.32 & 0.34 \\ 
  NE & 0.30 & 0.32 & 0.34 \\ 
  NH & 0.28 & 0.31 & 0.34 \\ 
  RI & 0.27 & 0.31 & 0.34 \\
\hline
\end{longtable}%
\end{minipage}
\label{statedist_modelpop_use2}
\end{table}

\newpage

\section{Summary of Individual-level Effects on Sharing Propensity}
\label{coef_summary}

\begin{table}[ht]
\centering
\caption{Summary table for the intercept covariate.} 
\label{summ_alpha}
%
\end{table}

\begin{table}[ht]
\centering
\caption{Summary table for the Sex:Female covariate.} 
%
\end{table}

\begin{table}[ht]
\centering
\caption{Summary table for the Party:Democrat covariate.} 
%
\end{table}

\begin{table}[ht]
\centering
\caption{Summary table for the Party:Neutral covariate.} 
%
\end{table}

\begin{table}[ht]
\centering
\caption{Summary table for the Party:Republican covariate.} 
%
\end{table}

\begin{table}[ht]
\centering
\caption{Summary table for the -18 age category covariate.} 
%
\end{table}

\begin{table}[ht]
\centering
\caption{Summary table for the 18-29 age category covariate.} 
%
\end{table}

\begin{table}[ht]
\centering
\caption{Summary table for the 30-39 age category covariate.} 
%
\end{table}

\begin{table}[ht]
\centering
\caption{Summary table for the 40+ age category covariate.} 
%
\end{table}

\begin{table}[ht]
\centering
\caption{Summary table for the state-wide fixed-effect coefficient on the  ``white\% population.} 
%
\end{table}

\begin{table}[ht]
\centering
\caption{Summary table for the state-wide educational attainment covariate.} 
%
\end{table}

\begin{table}[ht]
\centering
\caption{Summary table for the state population density covariate.} 
%
\end{table}

\clearpage

\section{Survey Questions}
\includepdf[pages=-]{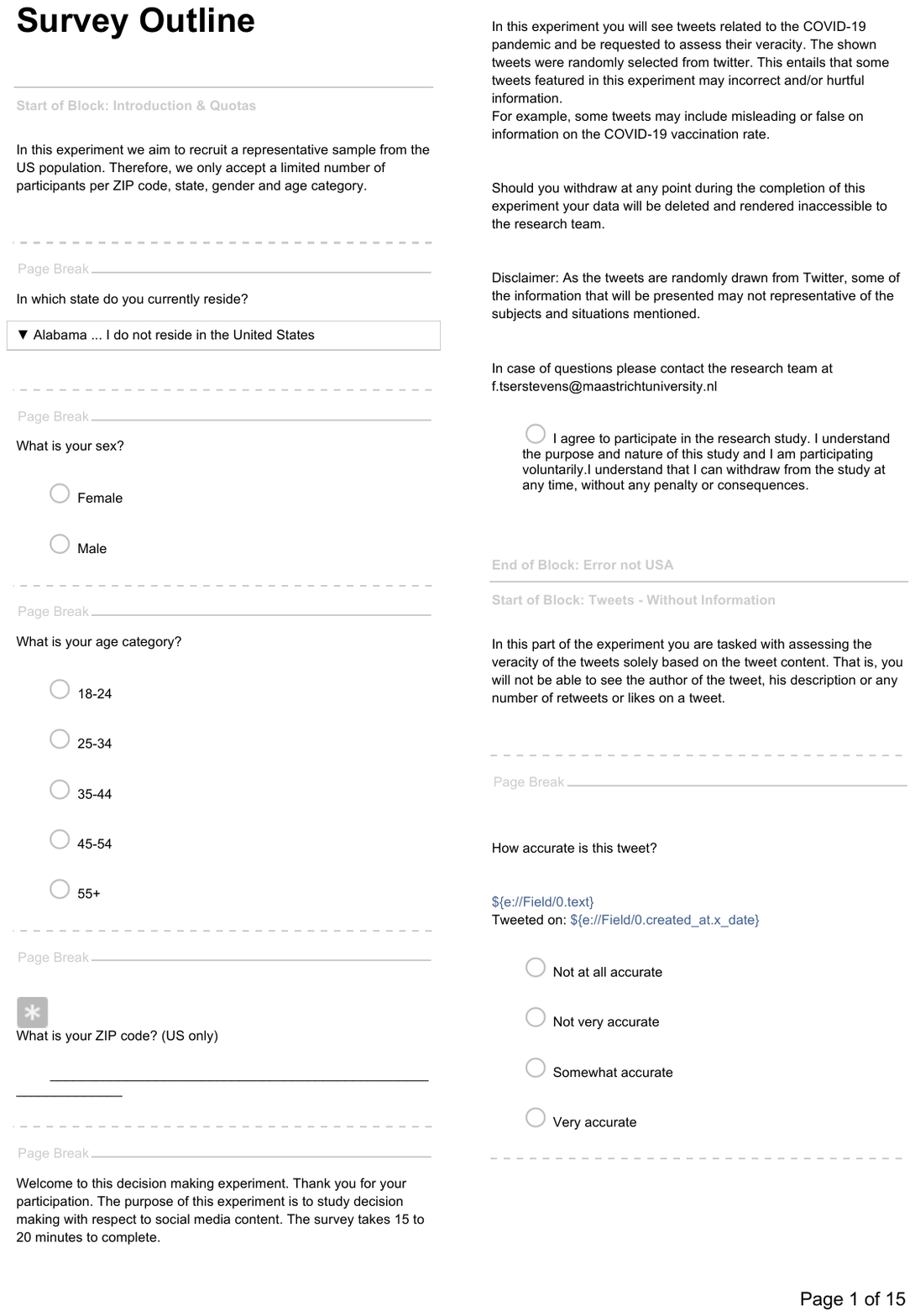}